\documentclass[twocolumn,superscriptaddress,amssymb,amsmath,footinbib,aps,prd,showpacs]{revtex4-1}
\pdfoutput=1

\usepackage{graphicx}
\usepackage{dcolumn}
\usepackage{hyperref}

\usepackage{amsfonts}
\usepackage{resizegather}
\usepackage{caption}
\usepackage{color}
\usepackage{subfigure}
\usepackage{ulem}
\usepackage{blindtext}

\usepackage{graphicx}

\begin{document}

\title{Graduated dark energy: Observational hints of a spontaneous sign switch in the cosmological constant}
\author{\"{O}zg\"{u}r Akarsu}
\email{akarsuo@itu.edu.tr}
\affiliation{Department of Physics, Istanbul Technical University, Maslak 34469 Istanbul, Turkey}

\author{John D. Barrow}
\email{J.D.Barrow@damtp.cam.ac.uk}
\affiliation{DAMTP, Centre for Mathematical Sciences, University of Cambridge, Wilberforce Road, Cambridge CB3 0WA, U.K.}

\author{Luis A. Escamilla}
\email{luis.escamilla@icf.unam.mx}
\affiliation{Instituto de Ciencias F\'isicas, Universidad Nacional Aut\'onoma de M\'exico, Cuernavaca, Morelos, 62210, M\'exico}

\author{J. Alberto Vazquez}
\email{javazquez@icf.unam.mx}
\affiliation{Instituto de Ciencias F\'isicas, Universidad Nacional Aut\'onoma de M\'exico, Cuernavaca, Morelos, 62210, M\'exico}
\begin{abstract}
We study the cosmological constant ($\Lambda$) in the standard $\Lambda$CDM model by introducing the \textit{graduated dark energy} (gDE) characterised by a minimal dynamical deviation from the null inertial mass density of the $\Lambda$ in the form $\rho_{\rm inert}\propto \rho^{\lambda}<0$ with $\lambda<1$ being a ratio of two odd integers, for which its energy density $\rho$ dynamically takes negative values in the finite past. For large negative values of $\lambda$, it creates a phenomenological model described by a smooth function that approximately describes the $\Lambda$ spontaneously switching sign in the late universe to become positive today. We confront the model with the latest combined observational data sets of PLK+BAO+SN+$H$. It is striking that the data predict bimodal posterior probability distributions for the parameters of the model along with large negative $\lambda$ values; the new maximum significantly excludes the $\Lambda$, and the old maximum contains the $\Lambda$. The improvement in the goodness of fit for the $\Lambda$ reaches highly significant levels, $\Delta\chi_{\rm min}^2=6.4$ for the new maxima, while it remains at insignificant levels, $\Delta\chi_{\rm min}^2\lesssim0.02$, for the old maxima. We show that, in contrast to the old maxima, which do not distinguish from the $\Lambda$, the new maxima agree with the model-independent $H_0$ measurements, high-precision Ly-$\alpha$ data, and model-independent $Omh^2$ diagnostic estimates. Our results provide strong hints of a spontaneous sign switch in the cosmological constant and lead us to conjecture that the universe has transitioned from AdS vacua to dS vacua, at a redshift $z\approx 2.32$ and triggered the late-time acceleration, and suggests looking for such mechanisms in string theory constructions.

\end{abstract}

\maketitle

\section{Introduction}
\label{Intro}

The standard Lambda Cold Dark Matter ($\Lambda$CDM) model, relying on the inflationary paradigm \cite{Starobinsky:1980te,Guth:1980zm,Linde:1981mu,Albrecht:1982wi}, has proven so far to be the most successful cosmological model that accounts for the dynamics and the large-scale structure of the universe. It is in excellent agreement with a wide variety of the currently available data \cite{Riess:1998cb,Ade:2015xua,Alam:2016hwk,Abbott:2017wau,Aghanim:2018eyx}. Nevertheless, in addition to its long standing profound theoretical issues relating to the $\Lambda$ (or conventional vacuum energy) \cite{Weinberg:1988cp,Sahni:1999gb,Peebles:2002gy,Padmanabhan:2002ji}, it has recently begun to suffer from persistent tensions of various degrees of significance between some existing data sets (see, e.g., \cite{tension02,tension03,Zhao:2017cud,Bullock:2017xww,Freedman:2017yms} for further reading). Such tensions are of great importance as detection of even small deviations from the standard $\Lambda$CDM model with high significance could have substantial implications on our understanding of the fundamental theories of physics underpinning it.

One of the most intriguing tensions reported so far is the significant deficiency in the Hubble constant $H_{0}$ value predicted by the cosmic microwave background (CMB) Planck data \cite{Ade:2015xua,Aghanim:2018eyx} using the base $\Lambda$CDM model when compared with the values by direct model-independent local measurements \cite{Riess:2016jrr,Riess:2018byc,Riess:2019cxk,Freedman:2019jwv}. The fact that it worsens for the simplest minimally coupled single-field quintessence models and is only partially relieved by phantom models (or quintom models) aggravates this tension as it suggests the elimination of these standard Dark Energy (DE) models \cite{Vagnozzi:2018jhn,DiValentino:2019exe,DiValentino:2019dzu} (see also \cite{Visinelli:2019qqu} for further references). Surprisingly, the situation changes if the DE energy density is not restricted to be strictly positive. It has been reported that a number of persistent low-redshift tensions, including the $H_0$ tension, may be alleviated by a dynamical DE whose energy density can assume negative values or vanish at a finite redshift \cite{Delubac:2014aqe,Aubourg:2014yra,Sahni:2014ooa,Mortsell:2018mfj,Poulin:2018zxs,Capozziello:2018jya,Wang:2018fng,Dutta:2018vmq,Banihashemi:2018oxo,Banihashemi:2018has,Visinelli:2019qqu,Ye:2020btb}.
 
The possible need for DE whose energy density can assume negative values was previously emphasised by the observation that, when the base $\Lambda$CDM model is considered, the Ly-$\alpha $ forest measurement of the baryon acoustic oscillations (BAO) by the BOSS collaboration prefers a smaller value of the dust density parameter than is preferred by the CMB data \cite{Delubac:2014aqe}. They reported a clear detection of DE consistent with $\Lambda>0 $ for $z<1$, but with a preference for a DE assuming negative energy density values for $z>1.6$ and argued that the Ly-$\alpha $ data from $z\approx 2.34$ can fit a non-monotonic evolution of $H(z)$, i.e., of the total energy density $\rho _{\mathrm{tot}}(z)$ --assuming general relativity (GR)-- which is difficult to achieve in any model with non-negative DE density \cite{Aubourg:2014yra}. In another study \cite{Sahni:2014ooa}, in line with this, it was argued that the Ly-$\alpha$ data can be accommodated by a physically motivated modified gravity model that alters $H(z)$ itself, and also that a further tension relevant to the Ly-$\alpha $ data can be alleviated in models in which $\Lambda$ is dynamically screened, implying an effective DE passing below zero and concurrently exhibiting a pole in its equation of state (EoS), at $z\sim 2.4$. DE models --either as a physical source or an effective source arising from a modified theory of gravity-- assumes negative energy density values have not been paid much attention so far (for reviews on DE and modified theories of gravity \cite{Copeland:2006wr,Caldwell:2009ix,Clifton:2011jh,DeFelice:2010aj,Capozziello:2011et,Nojiri:2017ncd,Nojiri:2010wj}). However, such scenarios are in fact familiar from an effective source (say, DE) defined by the collection of all modifications to the usual Einstein field equations in scalar-tensor theories, namely, when the cosmological gravitational coupling strength gets weaker with increasing redshift \cite{Boisseau:2000pr,Sahni:2006pa}. A range of other examples of effective sources crossing below zero also exist, including theories in which $\Lambda $ relaxes from a large initial value via an adjustment mechanism \cite{Dolgov:1982qq,Bauer:2010wj}, in cosmological models based on Gauss-Bonnet gravity \cite{Zhou:2009cy}, in braneworld models \cite{Sahni:2002dx,Brax:2003fv}, in loop quantum cosmology \cite{Ashtekar:2006wn,Ashtekar:2011ni}, in higher-dimensional cosmologies that accommodate dynamical reduction of the internal space \cite{Chodos:1979vk,Dereli:1982ar,Akarsu:2012vv,Russo:2018akp}, and generalisations of the form of the matter Lagrangian in a non-linear way  \cite{Akarsu:2017ohj,Board:2017ign,Akarsu:2019ygx}.

It is possible to seek such scenarios by following a minimalist approach, namely, starting with the minimal extensions to the standard $\Lambda$CDM model. The most natural one to consider is the addition of positive spatial curvature, e.g., that of the Friedmann-Robertson-Walker (FRW) spacetime which imitates a negative energy density source with an EoS parameter equal to $-1/3$. It is easy to check that, however, to screen $\Lambda$ at, e.g., $z\sim2.4$ for $\Omega_{\Lambda,0}\sim0.7$, its density parameter today is required to be $\Omega_{k,0}\sim-0.06$, which contradicts to the inflationary paradigm and is indeed not allowed, e.g., by the joint results of the recent Planck release \cite{Aghanim:2018eyx} suggesting spatial flatness to a $1\,\sigma$ accuracy of 0.2\%. If we stay loyal to the inflationary paradigm and then suppose flat space, the simplest source that can realise such a behaviour can be obtained by promoting the null \textit{inertial mass density} \cite{EllisRC,Ellis:1998ct} of the vacuum energy ($\rho_{\rm inert}=0$) to a negative constant, $\rho_{\rm inert}=\rm{const}<0$. The source $\rho_{\rm inert}=\rm{const}$ has recently been of interest to many as it mimics $\Lambda$ today while leading the universe to exhibit a future singularity dubbed as the Little Sibling of the Big Rip for $\rho_{\rm inert}={\rm const}<0$ and a finite future bounce for $\rho_{\rm inert}={\rm const}>0$ \cite{Bouhmadi-Lopez:2014cca,Bouali:2019whr}. However, in the light observational analyses carried out in this paper, $\rho_{\rm inert}=\rm{const}<0$ provides us with  neither a superior DE model w.r.t. the $\Lambda$, nor an improvement regarding the tensions of interest to us. For instance, the observational data suggest that its energy density changes sign at a redshift larger than 65 (i.e., when it is already negligible) and it is indistinguishable from $\Lambda$ today ($z\sim 0$), so clearly it cannot have consequences on the tensions we are concerned. The simplest next step may be to consider the minimum \textit{dynamical} deviation from the null inertial mass density, viz., in the form $\rho_{\rm inert}\propto \rho^{\lambda}<0$ with $\lambda$ being a real constant. The exponent $\lambda$ here will provide us with a more featured evolution of the energy density passing below zero at high redshifts. Importantly, for arbitrarily large negative values of $\lambda$, it resembles a step function in redshift describing a \textit{spontaneously sign switching cosmological constant} at a certain redshift. Accordingly, it can also be viewed as a phenomenological model described by a smooth function for approximately describing a vacuum energy that switches sign at a certain redshift and becomes positive just recently in the late universe and triggers the acceleration. A source having this form (but considering $\rho_{\rm inert}\propto \rho^{\lambda}>0$) was first suggested in \cite{Barrow:1990vx} (see also \cite{Barrow:1990nv,Barrow:1990td}) for introducing an intermediate inflationary scenario named \textit{graduated inflation}. It was physically motivated by the form of bulk viscous stresses in FRW models and their quantum counterparts when the bulk viscosity is proportional to a power of the density.

Accordingly, we shall call this source \textit{graduated dark energy} (gDE) as in this paper we study the present-day acceleration of the universe. In fact, more recently, it has also been considered as a DE (e.g., \cite{Nojiri:2004pf,Stefancic:2004kb,Stefancic:2005cs,Frampton:2011sp}). However, all these works focus on the future singularities and the  asymptotic dynamics of the universe by retaining the positivity of the energy density (the cases for the negative energy density are discussed only superficially). In contrast, here, we focus on its dynamics around the present time and utilise its sign-switching energy-density feature to address the tensions that arise within $\Lambda$CDM model when the data from the late universe are considered.

Such scenarios, in particular, the sign-switching cosmological constant that arises as a limiting case of the gDE, can be extremely appealing from a string theoretic perspective. Constructing metastable de Sitter (dS) vacua (provided by $\Lambda>0$) has notoriously been a challenging task in string theory and, so far, has not have been concretely achieved \cite{Maldacena:2000mw,Silverstein:2007ac,Danielsson:2009ff,Wrase:2010ew,Danielsson:2010bc,Danielsson:2011au,Chen:2011ac,Danielsson:2012et,Dasgupta:2014pma,Cicoli:2018kdo}. This has led many to suggest that string theory might not have any dS vacua at all \cite{Vafa:2005ui,Danielsson:2018ztv,Obied:2018sgi,Garg:2018reu,Ooguri:2018wrx,Palti:2019pca}. This would obviously have immense implications in cosmology and/or theoretical physics, as it seems to imply an inconsistency between string theory and the universe we live in \cite{Kachru:2003aw,Agrawal:2018own,Andriot:2018wzk,Colgain:2018wgk,Heisenberg:2018yae,Kinney:2018nny,Akrami:2018ylq,Murayama:2018lie,Han:2018yrk,Kinney:2018kew,Colgain:2019joh}. In contrast, an AdS (anti-de Sitter) background (provided by $\Lambda<0$) solution naturally arises in string theory or string theory motivated supergravities with broken/unbroken supersymmetry. Furthermore, the AdS space provides a very powerful setup to study various strongly coupled quantum field theories via the AdS/CFT (conformal field theory) correspondence \cite{Maldacena:1997re,Witten:1998qj}. Contrary to the case of dS, which can only arise with broken supersymmetry, there does seem to exist a large number of consistent AdS backgrounds that can be obtained from string theory. It has also recently been claimed that transition from AdS vacua to dS vacua could be realised in a noncommutative quantum field theory setup \cite{Franchino-Vinas:2019nqy}. Consequently, if we could show through gDE, that the observational data prefer a DE having $\rho\sim\rho_0>0$ (positive cosmological constant) for $z\sim0$ (just recently) and $\rho\sim-\rho_0<0$ (negative cosmological constant) for $z\gg0$ (most of the history of the universe), which realises at large negative $\lambda$ values of gDE, and that the persistent tensions arising within the standard $\Lambda$CDM model disappear/relax, this  would have far reaching implications for our understanding of the fundamental laws of physics. We will show, by means of gDE, that the observational data provide strong pointers in this direction. This leads us to conjecture that the cosmological constant has spontaneously switched sign and became positive, namely, the universe has transitioned from AdS vacua to dS vacua, at $z\sim 2.3$ and triggered the observed late-time acceleration, and we suggest looking for such mechanisms in string theory.

\section{Graduated dark energy}

The energy-momentum tensor describing an isotropic perfect fluid can be decomposed relative to a unique four-velocity, $u^\mu$, in the form, $T_{\mu\nu}=(\rho+p) u_\mu u_\nu+p g_{\mu\nu}$,
where $\rho$ is the relativistic energy density relative to $u^\mu$, $p$ is the isotropic pressure, $g_{\mu\nu}$ is the metric tensor, and $\nabla_{\nu}u^{\mu}u_{\mu}=0$ and $u_\mu u^{\mu}=-1$. The set of equations arise from the twice-contracted Bianchi identities, by Einstein field equations, $G_{\mu\nu} =-\kappa T_{\mu\nu}$, implies the conservation equations. Projecting parallel and orthogonal to $u_\mu$, we obtain the energy and momentum conservation equations, correspondingly,
\begin{eqnarray}
\label{eqn:constraint}
\dot\rho+\Theta\rho_{\rm inert}=0\quad\textnormal{and}\quad
{\rm D}^\mu p+\rho_{\rm inert}\dot{u}^\mu=0,
\end{eqnarray}
where $\rho_{\rm inert}=\rho+p$, the multiplier of the four acceleration $\dot{u}^\mu$, is \textit{the inertial mass density} \cite{EllisRC,Ellis:1998ct}. Here, ${\rm D}_\nu$ is the spatial gradient (the covariant derivative operator orthogonal to $u^\mu$) defined by ${\rm D}_\nu f= \nabla_\nu f+u_\mu\dot f$; $\Theta={\rm D}^\mu u_\mu$ is the volume expansion rate and overdots denote derivatives w.r.t. the comoving proper time $t$. 

Inspired by \cite{Barrow:1990vx}, we define a type of DE model, we named as \textit{graduated Dark Energy} (gDE), which yields an inertial mass density exhibiting power-law dependence to its energy density as follows;
\begin{equation}
\label{eqn:GDE}
\rho_{\rm inert}=\gamma\rho_0 \left(\frac{\rho}{\rho_0}\right)^{\lambda},
\end{equation}
where $\rho_0$ is positive definite (throughout the paper, subscript 0 attached to any quantity denotes its value today), the parameters $\gamma$ and $\lambda$ are real constants. This can be viewed as characterising the minimum dynamical deviation from the null inertial mass density, viz., from the conventional vacuum energy. So that equation of state (EoS) parameter is $w=p/\rho=-1+\rho_{\rm inert}/\rho$, and reads
\begin{equation}
w=-1+\gamma \left(\frac{\rho}{\rho_0}\right)^{\lambda-1}.
\end{equation}
We note that $\gamma=0$ corresponds to the conventional vacuum energy with $w=-1$ (leading to the $\Lambda$CDM model) and $\lambda=1$ corresponds to the perfect fluid with constant EoS parameter $w=-1+\gamma={\rm const}$ (leading to the $w$CDM model). From the continuity equation \eqref{eqn:constraint}, this leads to ${\rm d}\rho+3 \gamma\rho_0 \left(\frac{\rho}{\rho_0}\right)^{\lambda} \frac{{\rm d} {a}}{a}=0$, which is solved by
\begin{equation}
\label{eqn:rhogde}
\rho=\rho_0 \left[1+3\gamma  (\lambda-1) \ln a \right]^{\frac{1}{1-\lambda}},
\end{equation}
which satisfies
\begin{align}
\label{intertmd}
\rho_{\rm inert}&=\gamma \rho_0 \left[1+3 \gamma  (\lambda-1) \ln a \right]^{\frac{\lambda}{1-\lambda}},\\
w&=-1+\frac{\gamma }{1+3\gamma (\lambda-1) \ln a}.
\end{align}
We note that $w=-1+\gamma$ for today $a=1$ (redshift $z\equiv-1+\frac{1}{a}=0$), $w\approx-1$ for sufficiently large and small $a$, in particular, $w\rightarrow-1$ in the far future ($a\rightarrow\infty$) and in the very early universe ($a\rightarrow0$). Besides, provided that the parameters $\gamma$ and $\lambda$ are chosen appropriately, gDE can achieve transition from $\rho>0$ to $\rho<0$ at a certain redshift. Thus, gDE can also be viewed as a phenomenological model described by a smooth function for approximately describing the cosmological constant switches sign at a certain redshift and, for instance, becomes positive just recently in the late universe.

The gDE \eqref{eqn:rhogde}, in fact, exhibits various types of dynamics depending on its free parameters $\lambda$ and $\gamma$, see \cite{Stefancic:2004kb} for a comprehensive investigation. In this paper, we are interested in the case its energy density passes below zero at high redshifts, which, so far, has not been paid much attention, yet it is the case fitting the scenarios we discussed in the Introduction \ref{Intro} that most likely address the tensions relevant to $H_0$ and, in particular, to the high-precision Ly-$\alpha $ data from $z\approx 2.34$. For instance, in the case $\lambda=0$ ($\rho_{\rm inert}=\gamma\rho_0$), \eqref{eqn:rhogde} reduces to $\rho=\rho_0-3\rho_0\gamma\ln a $,  consisting of a constant $\rho_0>0$ mimicking $\Lambda>0$ and a dynamically screening term, $-3\rho_0\gamma\ln a $, in the past for $\gamma<0$, viz., $\rho_0-3\rho_0\gamma\ln a=0$ at $a=e^{\frac{1}{3\gamma}}$. Yet, the presence of the exponent $\frac{1}{1-\lambda}$ in \eqref{eqn:rhogde} will allow us to realise such a scenario with additional features.

First, we define $\rho/\rho_0=x^y$ along with $\rho_0>0$, where $x\equiv1+3\gamma  (\lambda-1) \ln a$ and $y\equiv\frac{1}{1-\lambda}$. We note that, unless $\gamma=0$ (conventional vacuum) or $\lambda=1$ (perfect fluid with constant EoS parameter), $x$ changes sign at
\begin{equation}
\label{eqn:sc}
a=a_*\equiv {\rm e}^{-\frac{1}{3}\frac{1}{\gamma(\lambda-1)}},
\end{equation}
which is in the past ($a_*<1$, the case we are interested in) for $\gamma(\lambda-1)>0$, and in the future ($a_*>1$) for $\gamma(\lambda-1)<0$. Next, $y<0$ for $\lambda>1$ so that $\rho\rightarrow\pm\infty$ as $a\rightarrow a_*$ and $y>0$ for $\lambda<1$ so that $\rho\rightarrow 0$ as $a\rightarrow a_*$, where the latter case is of interest to us. Thus, we proceed with the following two conditions serving our purpose:
\begin{equation}
\label{1stconditions}
\lambda<1\quad{\rm and}\quad \gamma<0,
\end{equation}
the latter of which implies $w(a=1)<-1$, i.e., the gDE must be in the phantom region today.

To get around a mathematical obstacle, when we investigate gDE computationally (see \cite{ourfootnote}), we continue by writing $\frac{\rho}{\rho_0}=x^y$ in an equivalent way as $\frac{\rho}{\rho_0}={\rm sgn}(x)\, {|x|}^y$ for $y=\frac{m}{n}$ with $m$ and $n$ being odd integers, namely,
\begin{equation}
\begin{aligned}
\label{eqn:greatDE}
\rho= \rho_0\, {\rm sgn}[1-\Psi \ln a] \,\big| 1-\Psi \ln a \big|^{\frac{1}{1-\lambda}},
\end{aligned}
\end{equation}
for $\Psi \equiv -3\gamma(\lambda-1)<0$ (i.e., $\gamma<0$), $\lambda<1$ and the exponent $\frac{1}{1-\lambda}=\frac{m}{n}$ with both $m$ and $n$ being odd integers. For practical reasons, we will consider $m=1$ and so $\lambda=-2N$ with $N=0,1,2,...$, i.e., $\lambda=0,-2, -4,...$ . Here ${\rm sgn}$ is the signum function that reads ${\rm sgn}(x)=-1,0,1$ for $x<0$, $x=0$ and $x>0$, respectively. Of course, in principle, there is an infinite number of such $\lambda$ values, not continuous, between the ones we listed above, and so we can treat $\lambda$ in \eqref{eqn:greatDE} as if it is continuous since one can always find an allowed $\lambda$ value indistinguishably close to a forbidden $\lambda$ value.

Consequently, the gDE-CDM model replaces the $\Lambda$ of the Friedmann equation of the standard $\Lambda$CDM model by the gDE \eqref{eqn:greatDE} serving our purposes and reads
\begin{equation}
\label{eqn:Hzsgn}
\frac{H^2}{H_0^2}=\Omega_{\rm r,0}a^{-4}+\Omega_{\rm m,0}a^{-3}+\Omega_{\rm DE,0}\,{\rm sgn}[1-\Psi\ln a]\, \big| 1-\Psi\ln a \big|^{\frac{1}{1-\lambda}},
\end{equation}
from which we also read off
\begin{equation}
\label{eqn:greatDEtocrit}
\frac{\rho_{\rm DE}}{\rho_{\rm c,0}}= \Omega_{\rm DE,0}\, {\rm sgn}[1-\Psi\ln a] \,\big| 1-\Psi \ln a \big|^{\frac{1}{1-\lambda}},
\end{equation}
where $\Psi<0$ and $\lambda=0,-2, -4,...$ (For further possibilities, see \eqref{eqn:greatDE} and the explanations following it.). Here, the subscripts r and m stand for relativistic source ($w_{\rm r}=\frac{1}{3}$) and dust matter ($w_{\rm m}=0$), respectively.

Regarding inertial mass density \eqref{intertmd}; when $\gamma<0$, if $1-\lambda$ is odd then $\lambda$ is even, and consequently we have the exponent $\frac{\lambda}{1-\lambda}=\frac{\rm [even]}{\rm [odd]}$ in \eqref{intertmd}, which in turn implies that $\rho_{\rm inert}\leq0$, that is, we can write
\begin{equation}
\label{intertmdabs}
\rho_{\rm inert}=\gamma \rho_0 \left|1+3 \gamma  (\lambda-1) \ln a \right|^{\frac{\lambda}{1-\lambda}},
\end{equation}
under the conditions derived above. It turns out that $\rho_{\rm inert}=0$ is the upper bound, viz., $\rho_{\rm inert,max}=0$.

We claimed above that gDE can also be viewed as a phenomenological model described by a smooth function that approximately describes the cosmological constant switching sign at a certain redshift and becoming positive just recently in the late universe. Indeed, under the conditions we consider, $\rho(a=1)>0$ and $\rho(a\ll a_*)/\rho(a\gg a_*)\approx -1$ along with $w(a\ll a_*)\approx w(a\gg a_*)\approx-1$, which imply that the energy density of the gDE at high redshifts not only passes below zero but also settles in a value almost equal to the negative of its present time value and remains almost there, say, all the way to the early times before which gDE is irrelevant to the dynamics of the universe anymore. Note that the EoS parameter is just slightly below (above) the phantom divide line for $a\gg a_*$ ($a\ll a_*$) with $a_*<1$, and $w\rightarrow-1$ only when either $a\rightarrow0$ or $a\rightarrow\infty$. Therefore, the energy density of gDE grows very slowly in the future and reaches arbitrarily large values in the very remote future, and also grows in negative values very slowly ---obviously, much slower than radiation and dust, both which then eventually dominate gDE in the finite past--- with the increasing redshift for $a\ll a_*$, and reaches arbitrarily large negative values in the beginning of the universe. We note, however, that for arbitrarily large negative values of $\lambda$, the energy density equation \eqref{eqn:greatDEtocrit} (or \eqref{eqn:greatDE}) transforms into a step function;
\begin{equation}
\label{eqn:step}
\frac{\rho_{\rm DE}}{\rho_{\rm c,0}}\rightarrow \Omega_{\rm DE,0}\, {\rm sgn}[1-\Psi\ln a]\quad\textnormal{as}\quad \lambda\rightarrow-\infty
\end{equation}
with an EoS parameter $w\rightarrow-1$. In this case, the energy density of gDE is non-dynamical except that it spontaneously changes sign at $a=a_*$. Thus, for large negative values of $\lambda$, gDE model is a very good approximation for describing a cosmological constant spontaneously switching sign at $z=z_*$, namely, in the limit $\lambda\rightarrow-\infty$, $\frac{\rho_{\rm DE}}{\rho_{\rm c,0}}=\Omega_{\rm DE,0}$ for $z<z_*$ and $\frac{\rho_{\rm DE}}{\rho_{\rm c,0}}=-\Omega_{\rm DE,0}$ for $z>z_*$.

\begin{figure}[t!]
\captionsetup{justification=raggedright,singlelinecheck=false,font=footnotesize}
\includegraphics[width=0.35\textwidth]{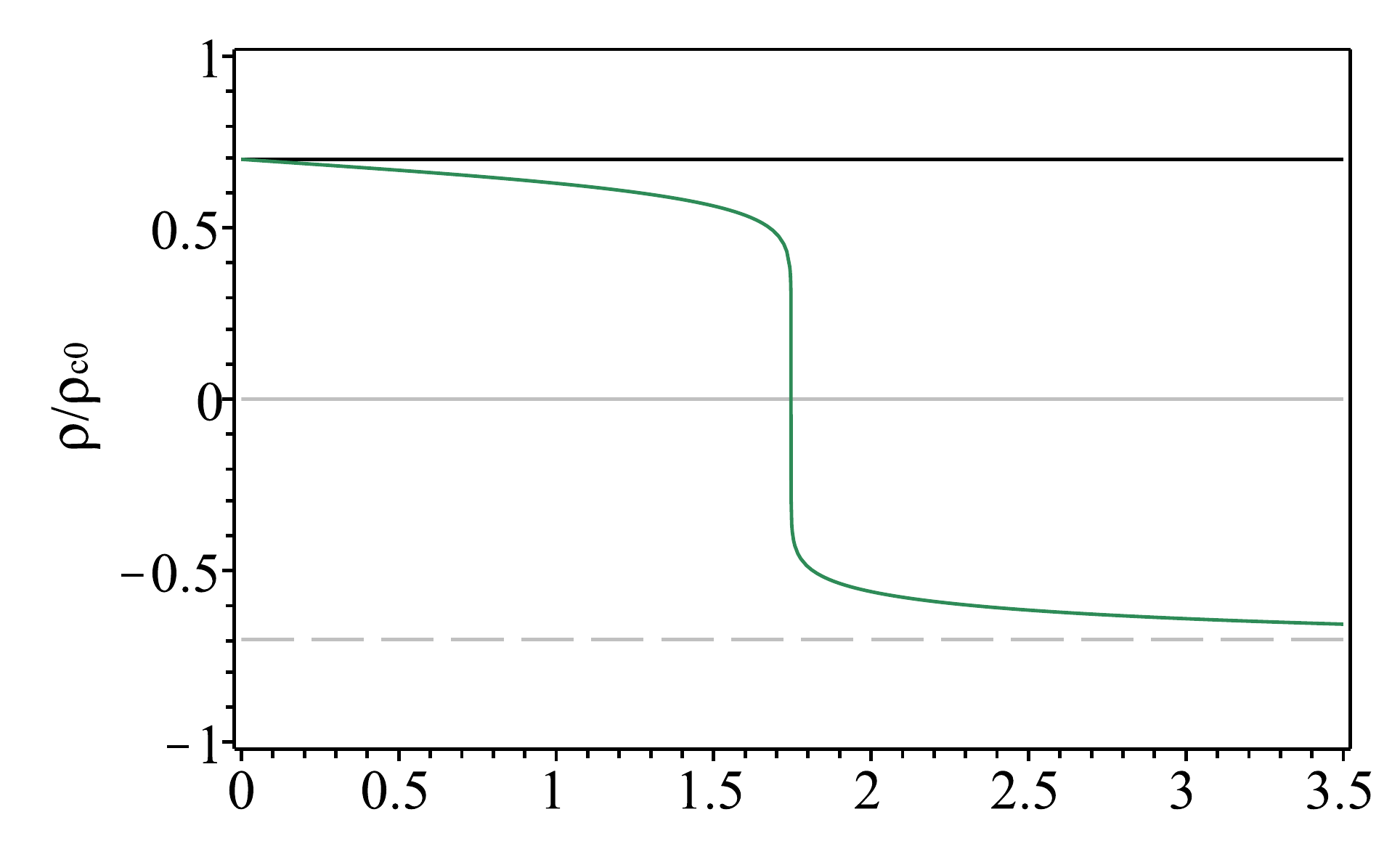}
\includegraphics[width=0.35\textwidth]{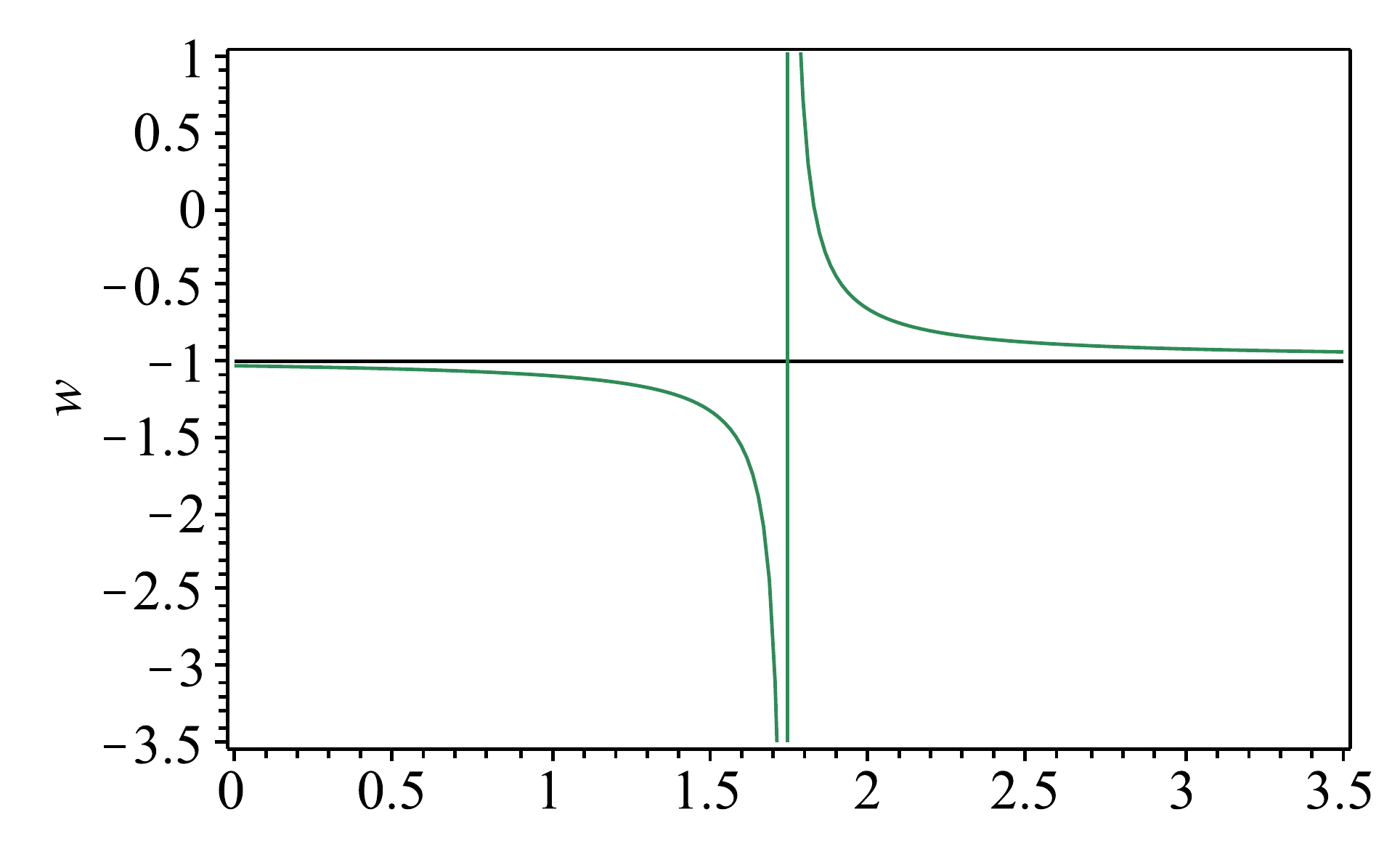}
\includegraphics[width=0.35\textwidth]{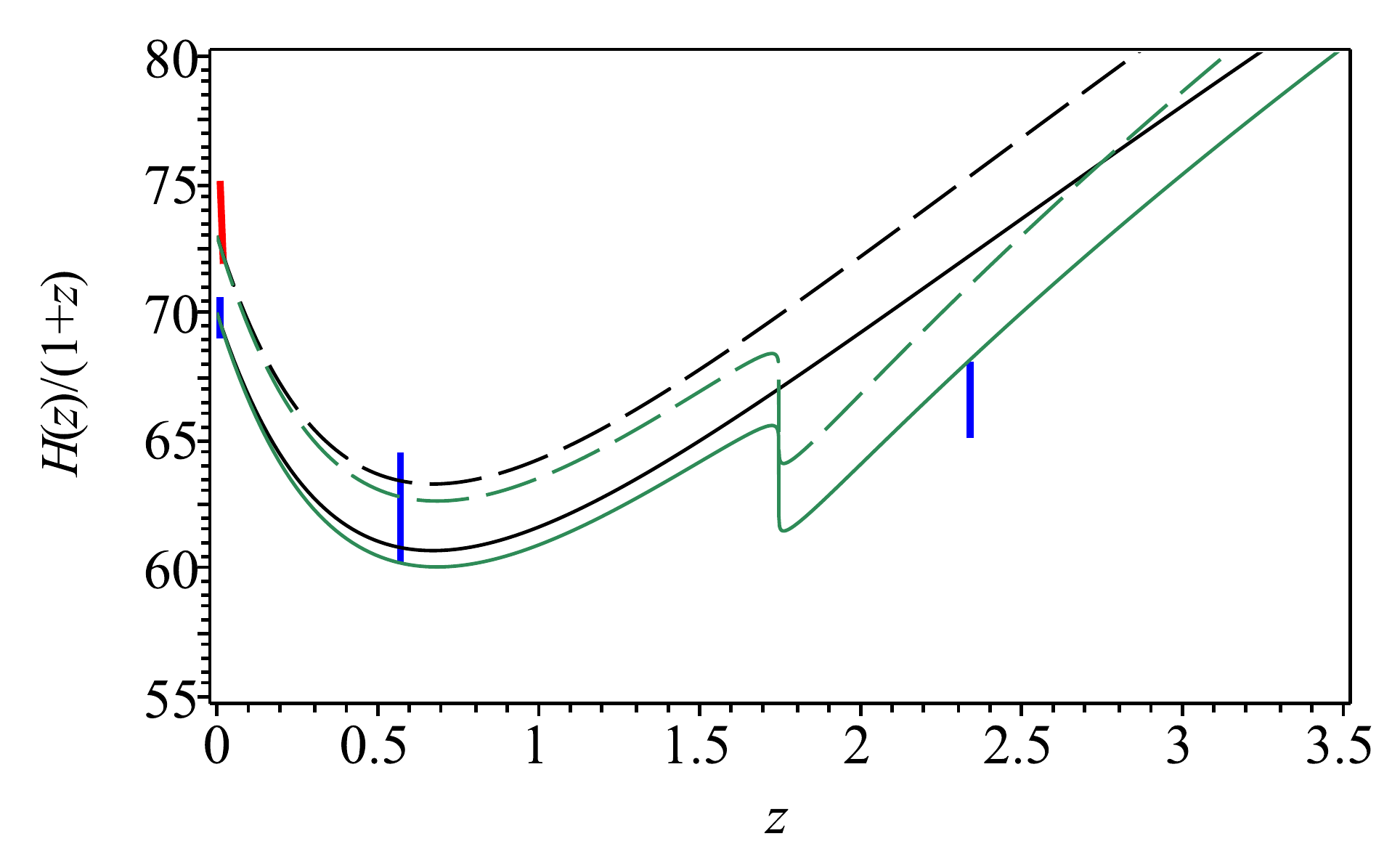}
\caption{We use $\Omega_{{\rm m},0}=0.30$ and, for gDE-CDM, $\gamma=-0.03$ along with $\lambda=-10$ (green). $H(z)/(1+z)$ vs. $z$ for the gDE-CDM (green) and $\Lambda$CDM (black). $H_{0}=70{\rm km\,s}^{-1}{\rm Mpc}^{-1}$ (solid) and $H_{0}=73{\rm km\,s}^{-1}{\rm Mpc}^{-1}$ (dashed). $H_0= 69.8 \pm 0.8\, {\rm km\,s}^{-1}{\rm Mpc}^{-1}$ from the TRGB $H_0$ \cite{Freedman:2019jwv}, $H(z=0.57) = 97.9 \pm 3.4\,{\rm km\,s}^{-1}{\rm Mpc}^{-1}$ \cite{Anderson:2013zyy}, and $H(z=2.34) = 222.4 \pm 5.0\,{\rm km\,s}^{-1}{\rm Mpc}^{-1}$ from the latest BAO data \cite{Delubac:2014aqe}. $H_0 = 73.52 \pm 1.62{\rm km\,s}^{-1}{\rm Mpc}^{-1}$ is independent measurement from Gaia parallaxes \cite{Riess:2018byc}.}
\label{trhosignchange}
\end{figure}

The following may be useful as a demonstration of how gDE-CDM model works and gives a guide to the values of the parameters of the model. Let us choose $a_*=e^{-1}$ ($z_*\sim 1.7$) in line with \cite{Aubourg:2014yra} (see Fig.11 in \cite{Aubourg:2014yra}). This leads to $\lambda=1+\frac{1}{3\gamma}$, where $\lambda$ must be a large negative number as we must use $\gamma\sim 0$ (it is observationally well known that $\gamma=w_0+1\sim0$) along with $\gamma<0$ (our condition derived above). For example, $\gamma=-0.03$ (or $w_0=-1.03$) predicted by the recent Planck release \cite{Aghanim:2018eyx} leads to $\lambda\sim-10$. Accordingly, in Fig.\ref{trhosignchange}, we depict $\frac{\rho(z)}{\rho_{\rm c,0}}$, $w(z)$ and  $H(z)/(1+z)$ by considering $\Omega_{\rm m,0}=0.30$ along with two different Hubble constant values, $H_0=70\, {\rm km\,s}^{-1}{\rm Mpc}^{-1}$ and $H_0=73\, {\rm km\,s}^{-1}{\rm Mpc}^{-1}$, for both the $\Lambda$CDM model and gDE-CDM model with $\lambda=-10$ and $\gamma=-0.03$. See the previous paragraph for the behaviours of $\rho$ and $w$ beyond our most interested redshift range $z=\{0,\,3.5\}$ considered in Fig. \ref{trhosignchange}. We note that, in the gDE-CDM model, the steep change in $H(z)/(1+z)$ at $z\sim z_*=1.7$ -- due to the sign change/pole of the energy density/EoS of the gDE-- allows it to pass through all data points as well as achieve larger $H_0$ values, whereas in the case of the $\Lambda$CDM model, it does not pass through Ly-$\alpha$ data at $z=2.34$ and the increased $H_0$ value worsens this situation. This is signalling that, w.r.t. the $\Lambda$, the gDE would lead to improved fit to the observational data and alleviate the tensions of various degrees of significance between some existing data sets within the $\Lambda$CDM cosmology. As, in the gDE-CDM model, we have $\rho\sim\rho_0$ and $w\lesssim-1$ (slightly in phantom region) for $z\ll z_*$ (also for $z\sim0$) and $\rho\sim-\rho_0$ and $w\gtrsim-1$ (slightly in quintessence region with negative energy density) for $z\gg z_*$, from phenomenological point of view such an achievement may be signalling that indeed the cosmological constant is responsible for the current acceleration of the universe, but it has changed sign at $z_*\sim 2$ and was negative at the higher redshifts.

\section{Constraints from the latest cosmological data}
\label{obsresults}

This section provides constraints on the gDE-CDM model using the latest observational data with a further discussion of the model and its consequences.

In order to perform the parameter-space exploration we implement a modified version of the simple and fast Markov Chain Monte Carlo code which computes expansion rates and distances from the Friedmann equation named SimpleMC \cite{Anze} and initially introduced in \cite{Aubourg:2014yra}. For a comprehensive review of the cosmological parameter inference see \cite{EPadilla}. The SimpleMC code takes into account a compressed version of recent datasets, for instance the Planck information (PLK) (where the CMB is treated as a ``BAO experiment" at redshift $z=1090$) measured by the angular scale of the sound horizon at that time, a recent analysis of Type Ia supernova (SN) data called Joint Light-curve Analysis compressed into a piece-wise linear function fit over 30 bins evenly spaced in $\log z$, and high-precision Baryon Acoustic Oscillation measurements (BAO), from comoving angular diameter distances, Hubble distance and the volume averaged distance, at different redshifts up to $z=2.36$. For a more detailed description about the datasets used see \cite{Aubourg:2014yra}. We also include a collection of currently available cosmic chronometer measurements ($H$), see \cite{Gomez-Valent:2018hwc}.

In this analysis, the radiation content is assumed by considering three neutrino species ($N_{\rm eff}=3.046$) with minimum allowed mass $\sum m_{\nu}=0.06\, {\rm eV}$ and a radiation density parameter given by $\Omega_{{\rm r},0}=2.469\times 10^{-5} h_0^{-2}(1+0.2271 N_{\rm eff})$, where $h_0$ is the present-day value of the dimensionless reduced Hubble parameter $h(z)=H(z)/100\, {\rm km\,s}^{-1}{\rm Mpc}^{-1}$ \cite{Dodelson03}. The total radiation content today is kept fixed in our analysis since it is well constrained by the CMB monopole temperature, $T_{{\rm CMB},0}=2.7255\pm 0.0006\,{\rm K}$ \cite{Fixsen09}.
Throughout our analysis we assume flat priors over our sampling parameters: $\Omega_{{\rm m},0}=[0.05,1.0]$ for the matter density parameter today, $\Omega_{{\rm b},0} h_0^2=[0.02,0.025]$ for the physical baryon density parameter and $h_0=[0.4,1.0]$ for the reduced Hubble constant. With regards to the gDE parameters, we assume $\gamma =[-0.2, 0]$ and $\lambda =[-27, 0]$ (when $\lambda$ is free). 

\begin{table*}[t!]
\captionsetup{justification=raggedright,singlelinecheck=false,font=footnotesize}
\footnotesize
\scalebox{0.9}{%
\begin{tabular}{cccccccc} 
\cline{1-8}\noalign{\smallskip}
 \vspace{0.15cm}

\qquad $\lambda$ \qquad&\qquad \qquad $\Omega_{\rm m,0}$ \qquad \qquad&\qquad \qquad $h_0$ \qquad \qquad&\quad \quad $\gamma=w_0+1$   \quad \quad&\qquad \qquad $\Psi$ \qquad \qquad&\qquad \quad $z_*$ \quad \qquad & \qquad $t_0$[Gyr]  \qquad &  $-2\Delta \ln \mathcal{L}_{\rm max}$  \\
\hline
\vspace{0.15cm}
$\Lambda$CDM & $0.302 (6)$  & $0.682 (5)$  & $0$  & 0 & $-$ & $13.806 (22)$ & 0.0\\
\vspace{0.15cm}
$0$ &  $0.297 (7)$ & $0.689 (7)$ & $> -0.08$  & $>-0.25$ & $-$ & $13.796 (24)$&  0.02 \\
\vspace{0.15cm}
$-2$&  $0.297 (7)$ & $0.688 (7)$ & $> -0.06$  & $>-0.61$ & $-$  & $13.795 (25)$ & 0.02 \\
 \vspace{0.15cm}
$-4$&  $0.289 (6)$, $0.298 (7)$  & $0.700 (9)$, $0.686 (7)$  & $-0.057 (2)$, $>-0.048$    
    &  $-0.86 (3)$, $>-0.73$ &  $2.31 (12)$,$-$  & $13.714 (25)$, $13.791 (26)$ & 1.0, 0.02  \\
\vspace{0.15cm}
$-6$&   $0.292 (6)$, $0.299 (6)$&  $0.699 (9)$, $0.685 (7)$  &  $-0.039 (1)$, $>-0.037$  
&   $-0.86 (3)$, $>-0.77$  & $2.31 (12)$,$-$ & $13.715 (25)$, $13.792 (27)$  &  2.0, 0.01  \\
 \vspace{0.15cm}
$-10$& $0.294 (6)$, $0.299 (6)$ & $0.696 (8)$, $0.684 (7)$ & $-0.025 (1)$, $>-0.021$ &
$-0.86 (3)$, $>-0.69$  & $2.32 (12)$,$-$ & $13.722 (27)$, $13.797 (25)$ & 4.4, 0.02  \\
 \vspace{0.15cm}
$-14$ & $0.296 (6)$, $0.300 (6)$ &  $0.695 (8)$, $0.683 (7)$  &  $-0.019 (1)$, $>-0.017$&
$-0.86 (3)$, $>-0.76$ & $2.33 (12)$,$-$ & $13.719 (31)$, $13.794 (27)$  &  5.3, 0.01   \\
 \vspace{0.15cm}
$-20$ & $0.297 (6)$, $0.300 (6)$ & $0.696 (9)$, $0.683 (7)$ &  $-0.013 (1)$, $>-0.012$ &
$-0.86 (3)$, $>-0.76$  & $2.32 (12)$,$-$ &  $13.718 (31)$, $13.795 (26)$ &6.0, 0.02  \\
 \vspace{0.15cm}
$-17.9 (5.8)$ & $0.296 (6)$, $0.299 (7)$ & $0.697 (9)$, $0.684 (8)$ & $-0.017 (8)$, $>-0.074$ 
& $-0.85 (4)$, $>-0.69$ & $2.32 (19)$,$-$ & $13.719 (30)$, $13.795 (24)$ & 6.4, 0.01  \\
\hline
\hline
\end{tabular}}
\caption{ Mean values along with $1\,\sigma$ constraints on the set of parameters used to described the gDE-CDM parameters. For one-tailed distributions the upper limit 
95\% C.L. is given. For two-tailed the 68\% C.L. is shown. 
The last column, $-2\ln (\mathcal{L}_{\Lambda{\rm,max}}/\mathcal{L}_{\rm gDE,max})$, is used to compute best-fit differences of gDE-CDM from $\Lambda$CDM ($-2\ln \mathcal{L}_{\Lambda, {\rm max}}=73.44$) based on the improvement in the fit alone.}
\label{TableI}
\end{table*}

\begin{figure}[t!]
\captionsetup{justification=raggedright,singlelinecheck=false,font=footnotesize}
    \centering
    \includegraphics[trim = 1mm  1mm 1mm 1mm, clip, width=4.2cm, height=4.cm]{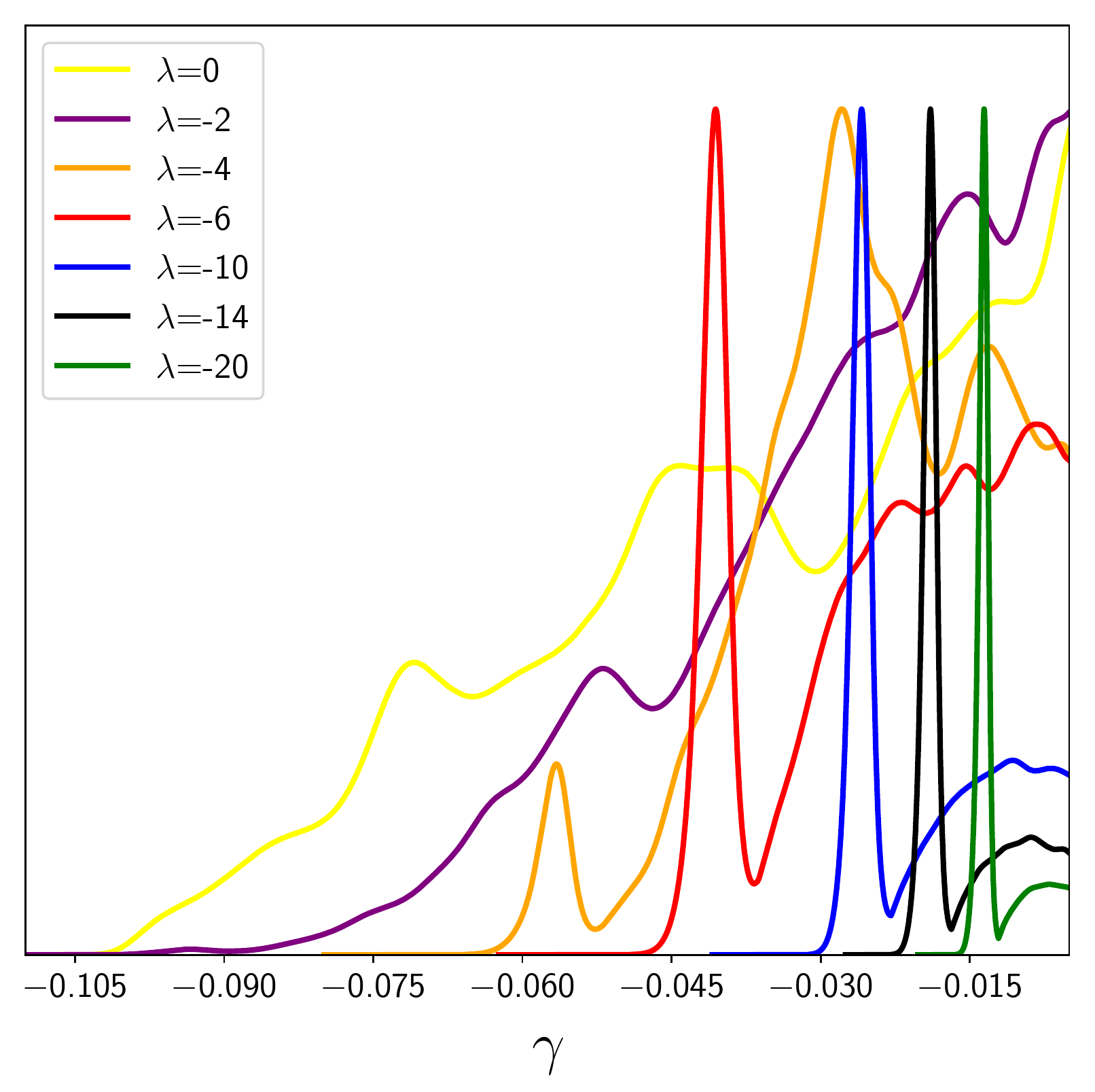}
    \includegraphics[trim = 1mm  1mm 1mm 1mm, clip, width=4.2cm, height=4.cm]{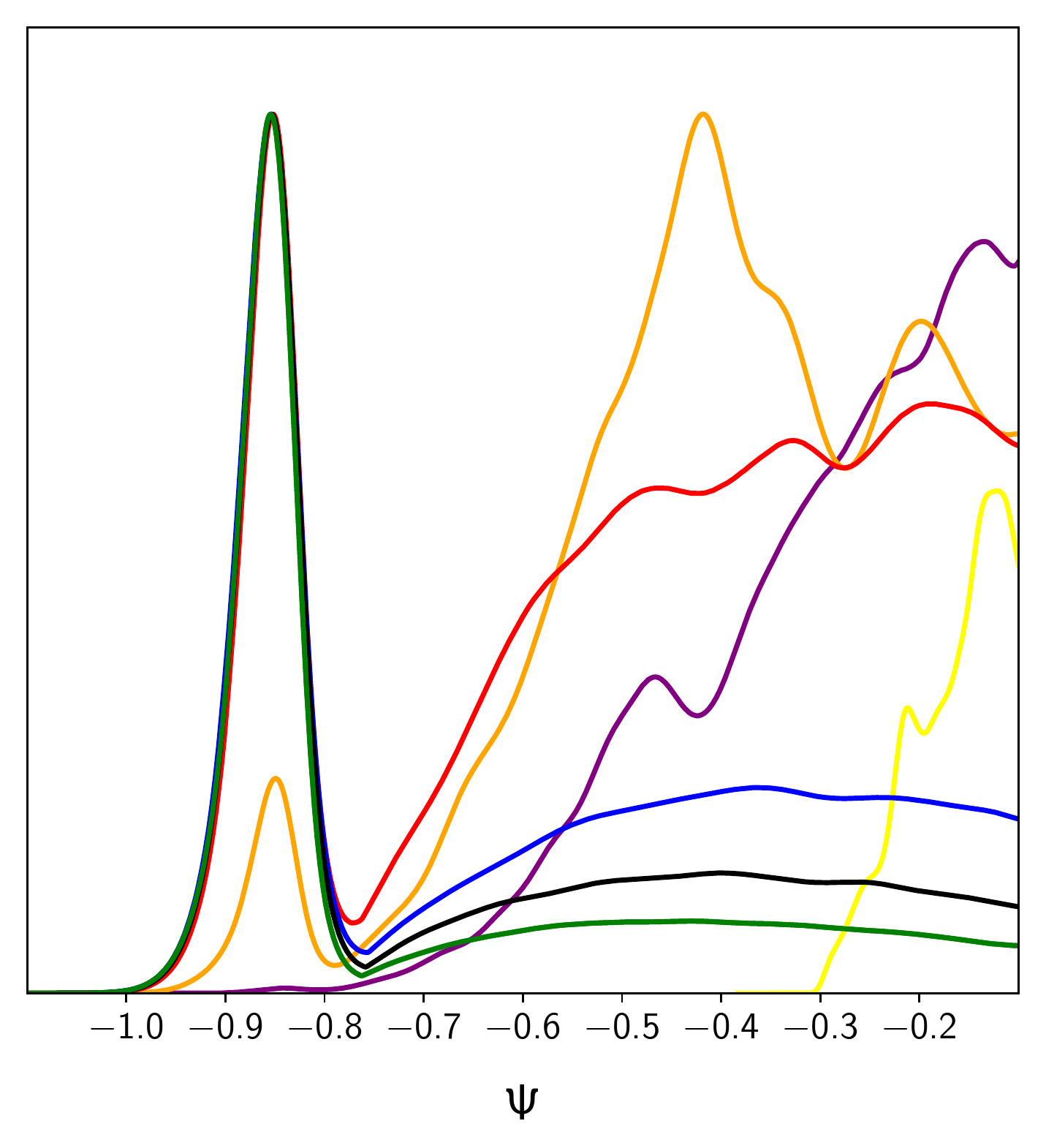}
     \includegraphics[trim = 1mm  1mm 1mm 1mm, clip, width=4.cm, height=4cm]{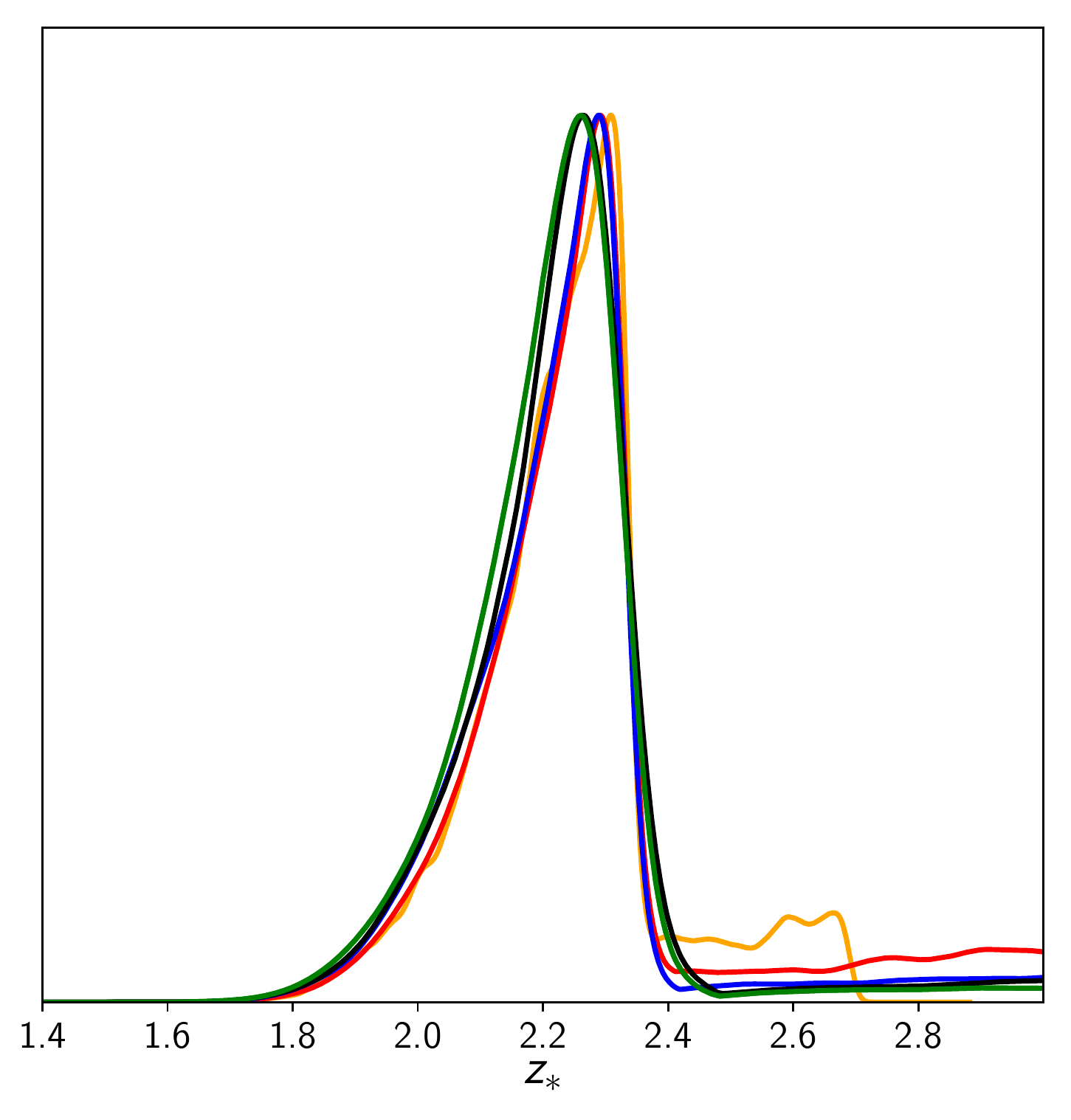}
    \caption{1D marginalised posterior distributions for the graduated $\gamma$ parameter (top left panel), 
        $\Psi \equiv 3\gamma(1-\lambda)$ (right) and the redshift location of the pole (if present) given by 
        Eqn. (\ref{eq:redshift}). For a better display we have included some particular cases of $\lambda$
        values. }
    \label{fig:lambda_1D}
\end{figure}

Table \ref{TableI} summarises the observational constraints on the free parameters --$\Omega_{\rm m,0}$, $h_0$, $\lambda$ and $\gamma$-- as well as the derived parameters --$\Psi$, $z_*$ and $t_0$ (age of the universe today)-- of the gDE-CDM model using the combined datasets PLK+BAO+SN+$H$; and for comparison shows those parameters used on the standard $\Lambda$CDM model ($\gamma=0$). The columns for each parameter contain the corresponding mean values and $1\,\sigma$ errors, according to the  number of modes presented on the 1D marginalised posterior distributions. In the last column we list the $-2\Delta \ln \mathcal{L_{\rm max}}=\Delta\chi_{\rm min}^2$ values representing the improvement in the fit to the data w.r.t. the $\Lambda$CDM. At the outset, we immediately notice that in our analyses the gDE leads to an improvement of up to $\Delta\chi_{\rm min}^2=6.4$ (corresponding to about $2.5\,\sigma$) w.r.t. the cosmological constant. In what follows we discuss in detail how this significant improvement is due to the fact that the gDE-CDM alleviates some of the tensions the $\Lambda$CDM experiences.

In Table \ref{TableI}, for $\lambda=0,-2$, we observe nothing interesting and no significant improvement to the fit w.r.t. $\Lambda$CDM, viz., $\Delta\chi_{\rm min}^2<0.02$. However, we observe something surprising occurs when $\lambda\leq-4$ (also when $\lambda$ is free) that the data predict bimodal posterior probability distributions for the parameters of the gDE-CDM, for which we observe two sets of constraint values in each column of Table \ref{TableI}. This may also be seen, for example, from the top left panel of Fig.~\ref{fig:lambda_1D} which displays 1D marginalised posterior distributions for the $\gamma$ parameters. Notice that, for $\lambda\leq -4$, as we move towards the larger negative values of $\gamma$, the existence of a second (new) maximum starts appearing significantly far away from $\gamma=0$ ($\Lambda$CDM). The first (old) maximum containing $\gamma=0$ is always there, but, when $\lambda\leq-6$, it consistently shrinks with the larger negative values of $\lambda$, during which the new maximum is getting relatively higher and sharper. This implies that the data significantly favour the new maximum over the old maximum when $\lambda\lesssim-6$. Indeed, we read from Table \ref{TableI} that the improvement in the fit w.r.t. $\Lambda$CDM reaches highly significant levels --e.g., $\Delta\chi_{\rm min}^2=6$ when $\lambda=-20$ and $\Delta\chi_{\rm min}^2=6.4$ when $\lambda$ is free-- for the new maximum, while it remains always at insignificant levels --$\Delta\chi_{\rm min}^2\lesssim0.02$ irrespective of the value of $\lambda$-- for the old maximum. The poor improvement level of $\Delta\chi_{\rm min}^2\lesssim0.02$ both in the old maximum (the maximum containing $\gamma=0$ when $\lambda\lesssim-4$ and $\lambda$ is free, and the single maximum when $\lambda\lesssim3$) presents no evidence for favouring these over the $\Lambda$CDM and the constraints on the parameters for these cases do not show a considerable deviation from those of the $\Lambda$CDM. Therefore, in what follows we discard all these cases and proceed our discussions  with reference to the $\Lambda$CDM ($\gamma=0$), basically, by considering only the new maximum that appears when $\lambda\lesssim-6$, e.g., by considering the one on the left of the pair of constraints given in a column for a parameter of the gDE-CDM in Table \ref{TableI}.

The presence of these new maxima has important consequences and may be better explained through the expression \eqref{eqn:sc}. This expression indicates if there exists a sign change in the energy density of the gDE (or a pole in its EoS parameter), it will happen at a redshift 
\begin{equation}\label{eq:redshift}
    z_*={\rm e}^{-\frac{1}{\Psi}}-1.
\end{equation}
Hence, the quantity $\Psi = -3\gamma(\lambda-1)$ determines the position of the pole and, if it is a real one, must yield a unique value irrespective of the values $\lambda$ and $\gamma$. That is, for a given $\lambda$, the $\gamma$ parameter selects its best position such that $\Psi$ remains unchanged, and this can be seen in the right-hand panel of Fig.~\ref{fig:lambda_1D} (see also Table \ref{TableI}). We observe that a peak at $\Psi=-0.86$ --significantly away from $\Psi=0$ ($\Lambda$CDM)-- emerges when $\lambda=-4$ and as $\lambda$ takes more negative values (see the cases $\lambda\leq-6$) it becomes significantly higher and sharper, fixed at $\Psi=-0.86$, while the old peak containing $\Psi=0$ becomes more prolate and lower. This implies highly significant observational evidence for the sign change of the energy density of the gDE (or pole in its EoS parameter) at the redshift corresponding to $\Psi=-0.86$.  We have shown, according to \eqref{eq:redshift}, in the bottom panel of Fig. \ref{fig:lambda_1D}, the 1D marginalised posterior distribution of the redshift for this event persistently located at $z_*\approx2.32$ (see Table \ref{TableI}). Interestingly, but not surprisingly this particular position agrees with the location of the Ly-$\alpha$ auto and cross-correlation BAO ($z=2.34$) data and the works \cite{Delubac:2014aqe,Aubourg:2014yra,Sahni:2014ooa}. This suggests such a behaviour of DE for alleviating the tensions besetting this observation. We should note here that the peaks containing $\Psi=0$ ($\Lambda$CDM)  also predict the sign change of the gDE, but we have discarded them for the following reasons. Firstly, these cases correspond to the ones we have discarded above, since they do not present any statistical evidence for being favoured over $\Lambda$CDM (the $\Psi\rightarrow0$ limit leading to $z_*\rightarrow\infty$). Secondly, in our analyses, we observe that these cases predict completely different $z_*$ values for different $\lambda$ values (if they were real the predictions  need to have been stable at a certain redshift) and all of which are extremely large (even having redshift values larger than the redshift of the big bang nucleosynthesis epoch) at which dark energy is irrelevant to the cosmological dynamics.
 
\begin{figure*}
\captionsetup{justification=raggedright,singlelinecheck=false,font=footnotesize}
\centering
\begin{center}     
\subfigure{\includegraphics[width=3.5cm,height=5.5cm,trim={0 0 0 0},clip]{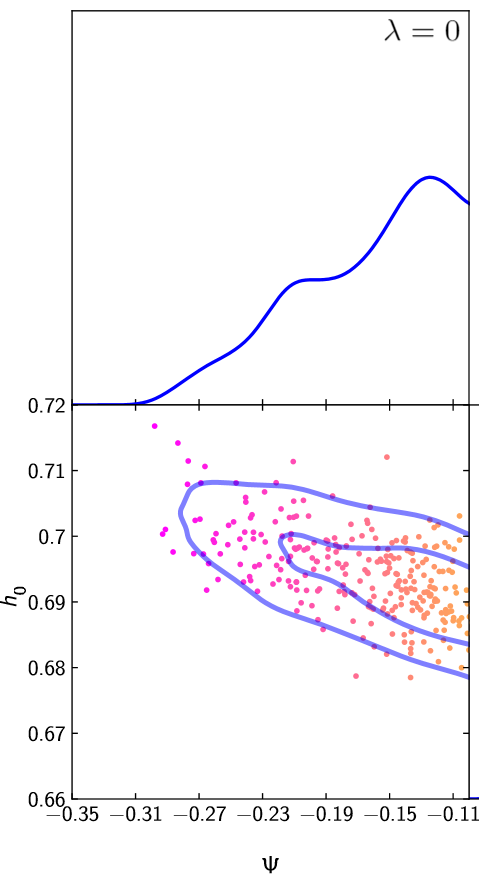}}%
\subfigure{\includegraphics[width=3.5cm,height=5.5cm,trim={0 0 0 0},clip]{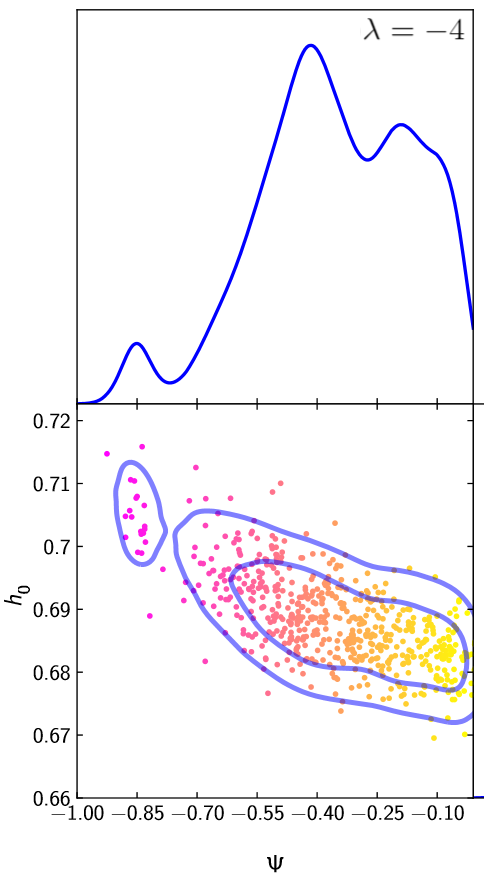}}%
\subfigure{\label{fig:a}\includegraphics[width=3.5cm,height=5.5cm,trim={0 0 0 0},clip]{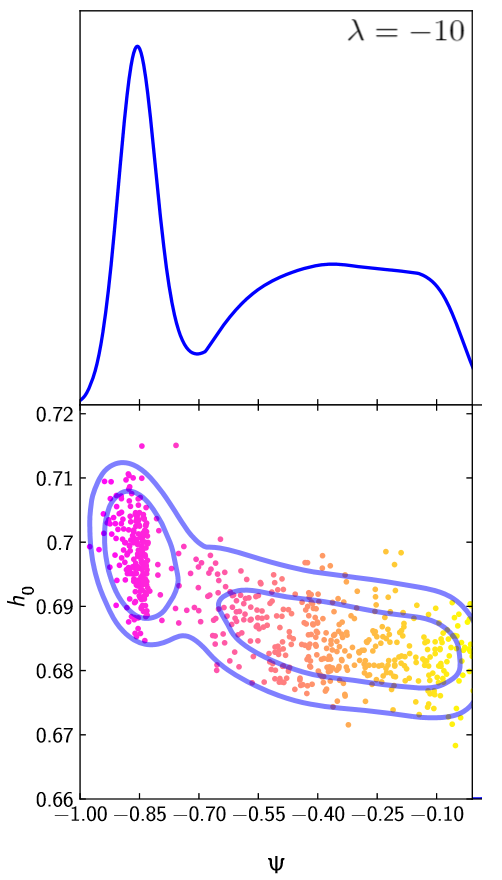}}%
\subfigure{\includegraphics[width=3.5cm,height=5.5cm,trim={0 0 0 0},clip]{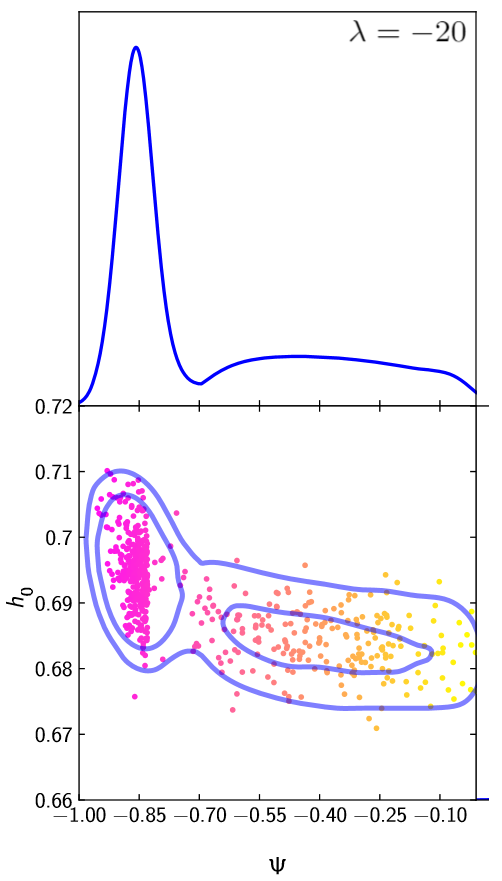}}%
\subfigure{\includegraphics[width=3.5cm,height=5.5cm,trim={0 0 0 0},clip]{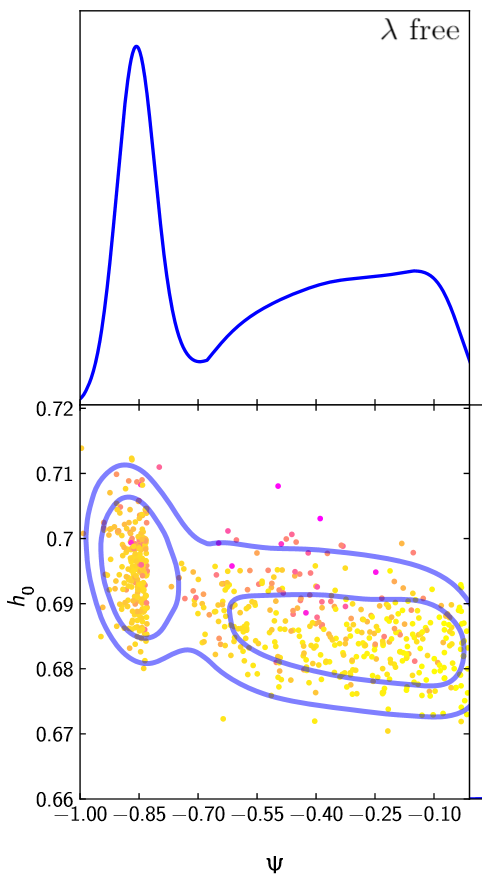}}%
\caption{Top panel: 1D marginalised posterior distributions of $\Psi$, along with (bottom panel)
2D posterior distributions of \{$\Psi$, $h_0$\} colour coded by the $\gamma$ parameter.}
\label{fig:psi_lam}
\end{center}
\end{figure*}
 
\begin{figure}[t!]
\captionsetup{justification=raggedright,singlelinecheck=false,font=footnotesize}
    \centering
    \includegraphics[scale=0.77]{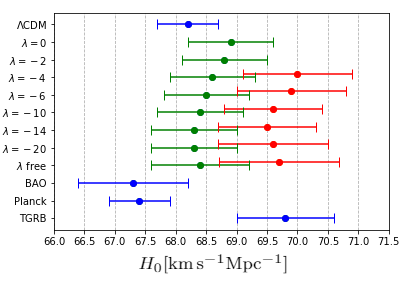}
    \caption{Means values along with $1\,\sigma$ error bars from the 1D marginalised posterior distributions of
    $H_0 [{\rm km\,s}^{-1}{\rm Mpc}^{-1}]$. Green error bars are associated with the peak 
    containing $\Psi\sim 0$ ($\Lambda$CDM), whereas  red with the new peak stable at $\Psi\sim -0.86$.} 
    \label{fig:my_h}
\end{figure}

The bimodal distribution that $\Psi$ exhibits has a strong impact on the posterior distribution of $h_0$, and therefore on the Hubble constant $H_0$, which also exhibits a bimodal behaviour. Fig.~\ref{fig:psi_lam} describes this behaviour; as soon as the $\lambda$ parameter starts decreasing the bimodal distribution on the panel $\{h_0,\Psi\}$ starts showing up for a particular $\gamma$ value (display in pink colour). This bimodal distribution is summarised on the marginalised error bars shown in Fig.~\ref{fig:my_h}. We observe that while the values (green) associated with the old peak containing $\Psi\sim 0$ ($\Lambda$CDM) agree with the $H_0$ values measured from the inverse distance ladder (e.g., $H_0= 67.4\pm.5$ from Planck 2018 \cite{Aghanim:2018eyx}), the ones (red) associated with the new peak stable at $\Psi\sim -0.86$ (away from $\Psi=0$) agree with the higher $H_0$ values measured from the distance ladder measurements (e.g., $H_0= 69.8\pm 0.8$ from a recent calibration of the Tip of the Red Giant Branch (TRGB) applied to Type Ia supernovae \cite{Freedman:2019jwv}). Therefore, the $H_0$ predicted within the $\Lambda$CDM (matching our results from the old peak) has  deficiency w.r.t. the TRGB $H_0$ value, while the ones predicted by the new peak (appears for $\lambda\lesssim-4$) perfectly match with it. It certainly favours the new peak that it predicts a value matching the independent TRGB $H_0$ value. It is also significant that it uses the distance ladder approach, rather than the inverse distance ladder approach. Also, the latter BAO calibration of $H_0$ is not completely independent of the Planck measurement, as both $H_0$ determinations are based on the $\Lambda$CDM and its adopted value of the sound horizon scale. Moreover, the independent TRGB $H_0$ value (so the values from our new peak) agrees with both Planck \cite{Aghanim:2018eyx} and Cepheid \cite{Riess:2016jrr,Riess:2018byc,Riess:2019cxk} $H_0$ values. However,  when combined with Cepheid measurements the tension with the Planck value is relieved only at about $\sim 1\sigma$ level and still remains significant \cite{Freedman:2019jwv}.

\begin{figure}[!t]
\captionsetup{justification=raggedright,singlelinecheck=false,font=footnotesize}
\centering
\includegraphics[trim = 3mm  1mm 0mm 1mm, clip, width=5.0cm, height=4.cm]{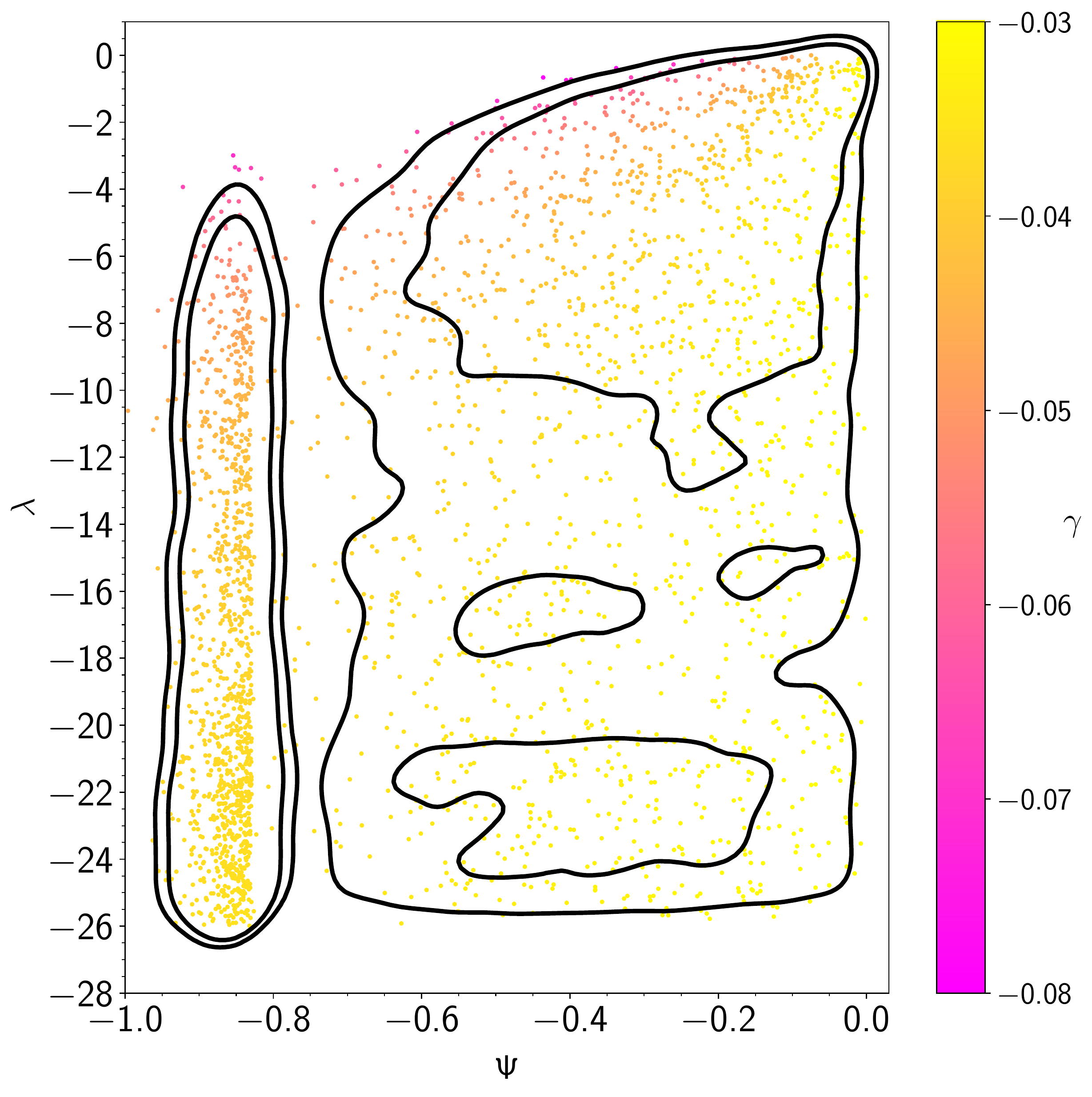}
\includegraphics[trim = 2mm  1mm 2mm 1mm, clip, width=3.5cm, height=4cm]{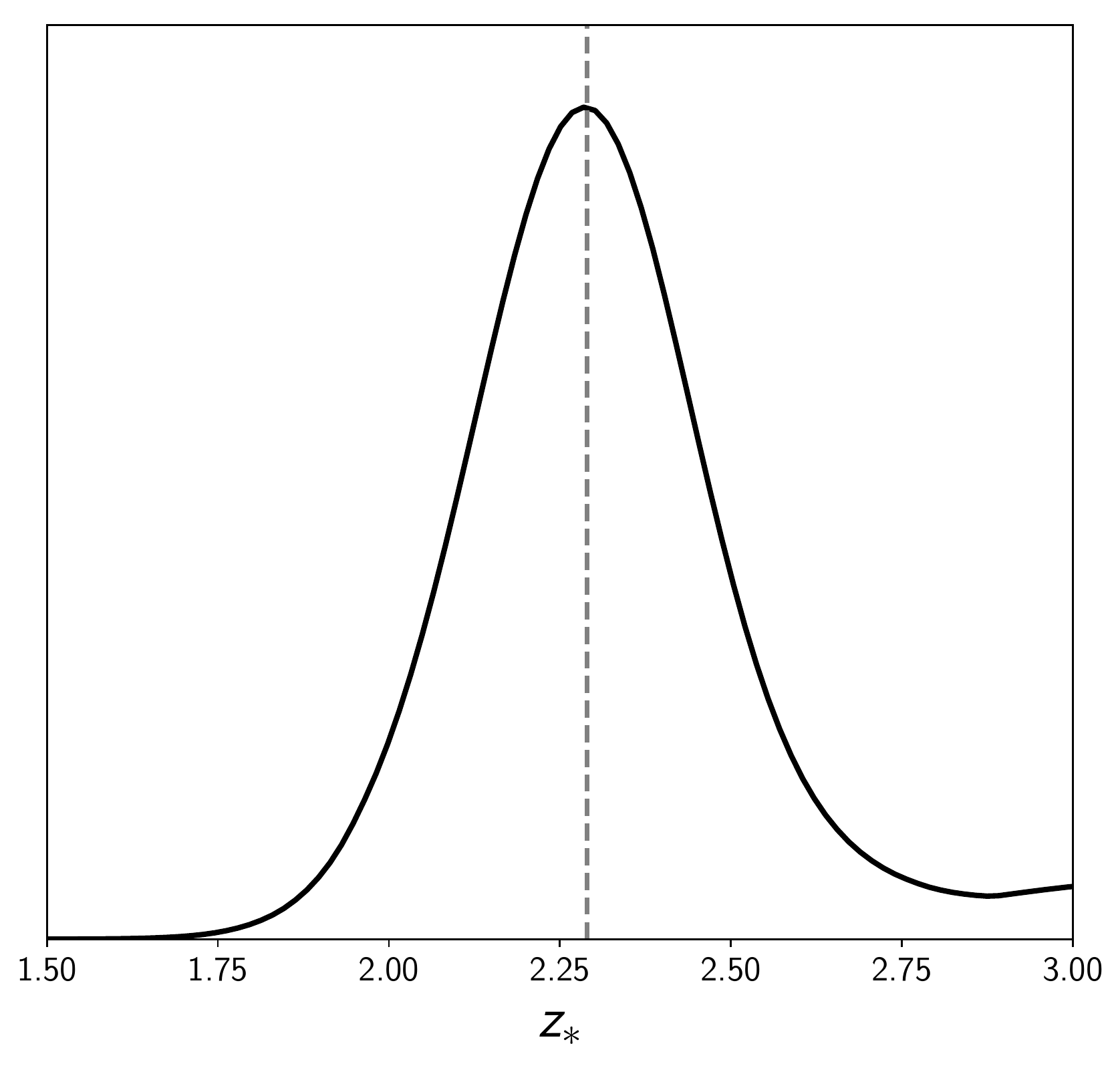}
\caption{Graduated Dark energy model with varying the $\lambda$ parameter. Left panel: 3D marginalised posterior distributions for the graduated $\lambda$ and $\Psi$ parameters, 
    coloured coded by the $\gamma$ parameter. Right panel: 1D marginalised posterior of the redshift position 
    given by the pole. The vertical line is the mean value $z_*=2.32$.}
\label{fig:lambda_2D}
\end{figure}

 We notice in Table \ref{TableI} that the values of the parameters $\Psi(\gamma,\lambda)$ --or $z_*(\gamma,\lambda)$-- and of the other cosmological parameters $\Omega_0$, $h_0$ and $t_0$ are quite stable for $\lambda\leq-10$. One may see from the last row in Table \ref{TableI} that we confirm this observation when we constrain the model by letting also the parameter $\lambda$ free (we use flat prior $\lambda= [-27, 0]$). Left panel of Fig.~\ref{fig:lambda_2D} displays the 3D marginalised posterior distribution of the $\{\Psi$,$\lambda\}$ parameter region colour coded with the $\gamma$ parameter. Here, the bimodality of the constraints on the gDE-CDM shows up as two detached 2D outer contours. The narrow one located at $\Psi\sim -0.86$ corresponds to the new maximum, while the wide one corresponds to the old maximum containing the $\Lambda$CDM (top-right corner). In the right panel of the same figure we present the 1D posterior distribution of the $z_*$ associated with the new maximum, which demonstrates that the redshift at which the gDE energy density changes sign (its EoS parameter exhibits a pole) is stable at $z_*\sim 2.32$.
 
 It was shown in \cite{Sahni:2014ooa} through the $Omh^2$ diagnostic (introduced to test the $\Lambda$ hypothesis in a model-independent way) that the $\Lambda$CDM is in tension with the BAO's statistically independent measurements of $H(z)$ at redshifts of 0.57 and 2.34. It was shown that this tension is alleviated in models in which the $\Lambda$ was dynamically screened (compensated) in the past and that the energy density of such evolving DE models passes below zero (exhibits pole in the effective EoS) at $z\sim 2.4$. These are in line with the new maxima of the gDE-CDM, yet in addition the fact that the constant that plays the role of $\Lambda$ in gDE is embedded into a parenthesis raised to a power renders our model more featured. Therefore, we also investigate gDE in the context of $Omh^2$ diagnostic.
 
  The $Omh^2$ diagnostic is defined in \cite{Sahni:2014ooa} as follows:
\begin{equation} \label{def:Omh2}
Om h^2(z_i ; z_j)=\frac{h^2(z_i)-h^2(z_j)}{(1+z_i)^3-(1+z_j)^3},
\end{equation} 
 and depends only on $H(z)$. Accordingly, knowing it at two or more redshifts, one can obtain $Omh^2$ value(s) in a model-independent manner and thence conclude whether or not the DE is a $\Lambda$. For the $\Lambda$CDM, omitting radiation (negligible in the late universe), we have $h^2=h_0^2 \left[\Omega_{\rm m,0} (1+z)^3+1-\Omega_{\rm m,0}\right]$ leading to a constant
\begin{equation}
Om h^2(z_i ; z_j)=h_0^2 \Omega_{\rm m,0}.
\end{equation}
For the gDE-CDM, using \eqref{eqn:Hzsgn}, we have
\begin{equation}
\begin{aligned}
&Om h^2(z_i ; z_j)=h_0^2 \Omega_{\rm m,0}\\
&\quad\quad+h_0^2\,(1-\Omega_{\rm m,0})\frac{{\rm sgn}(x_i)|x_i|^y-{\rm sgn}(x_j)|x_j|^y}{(1+z_i)^3-(1+z_j)^3},
\end{aligned}
\end{equation}
where have neglected radiation and used the zero-curvature constraint, $\Omega_{\rm m,0}+\Omega_{\rm DE,0}=1$. The second line of the $Om h^2(z_i ; z_j)$ for the gDE-CDM emerges as a correction to the one for the $\Lambda$CDM. We can calculate  the predicted $Om h^2(z_i ; z_j)$ with these two equations for any pair of chosen redshifts using the constraints on the models and then compare the same with the model-independent estimates obtained by \eqref{def:Omh2}.

\begin{table}
\captionsetup{justification=raggedright,singlelinecheck=false,font=footnotesize}
\footnotesize
\scalebox{0.9}{%
\begin{tabular}{cccc} 
\cline{1-4}\noalign{\smallskip}
\vspace{0.15cm}
$\lambda$ & $\quad Omh^2(z_1;z_2)\quad$ & $\quad Omh^2(z_1;z_3)\quad$ & $\quad Omh^2(z_2;z_3)\quad$ \\
\hline
\vspace{0.15cm}
$\Lambda$CDM       &  $0.140 (2)$  	&  $0.140(2)$     &         $0.140 (2)$ \\
\vspace{0.15cm}
0        &  $0.134 (4)$ 		& $0.139 (4)$   &  $0.140 (4)$ \\
\vspace{0.15cm}
-2        &  $0.135 (4)$ 		& $0.140 (2)$   &  $0.140 (2)$ \\
\vspace{0.15cm}
-4        &  $0.136 (3)$ 		& $0.129 (1)$, $0.140 (2)$  &  $0.129 (2)$, $0.140 (2)$\\
\vspace{0.15cm}
-6        &  $0.137 (2)$ 		& $0.128 (1)$, $0.140 (3)$  &  $0.127 (2)$, $0.140 (2)$\\
\vspace{0.15cm}
-10       &  $0.137 (2)$, $0.139 (2)$    &   $0.127 (2)$, $0.140 (2)$   & $0.123 (2)$, $0.140 (2)$\\
\vspace{0.15cm}
-14       &   $0.138 (2)$, $0.139 (2)$	 & $0.126 (2)$, $0.140 (2)$  &  $0.127 (2)$, $0.140 (2)$  \\
\vspace{0.15cm}
-20       &   $0.139 (2)$, $0.140 (2)$ & $0.125 (2)$, $0.140 (2)$  &  $0.124 (2)$, $0.140 (2)$  \\
\vspace{0.15cm}
Free      &      $0.136 (4)$, $0.139 (2)$	 &   $0.127 (4)$, $0.140 (2)$ &  $0.126 (2)$, $0.140 (2)$  \\
\hline
\hline
\end{tabular}}
\caption{Mean values along with 1$-\sigma$ constraints on the set of parameters that describe $Om$ diagnostic.}
\end{table}

\begin{figure}[t!]
\captionsetup{justification=raggedright,singlelinecheck=false,font=footnotesize}
    \centering
\subfigure{\label{fig:a}\includegraphics[trim={0 0 4cm 0.90cm},clip, width=3.cm, height=4.cm]{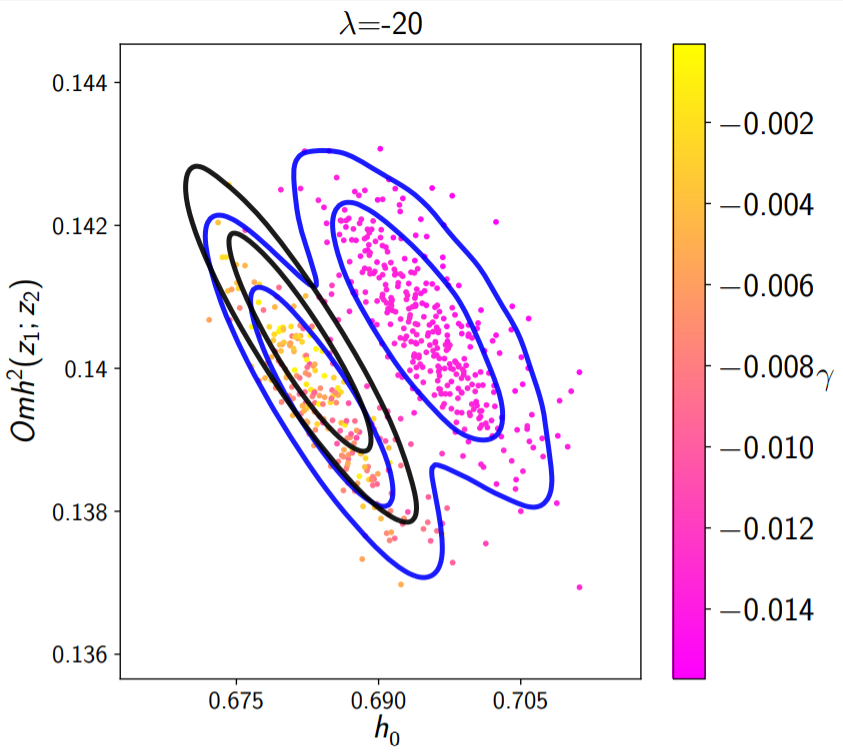}}%
\subfigure{\label{fig:b}\includegraphics[trim={0 0 4cm 0},clip, width=3.cm, height=4.2cm]{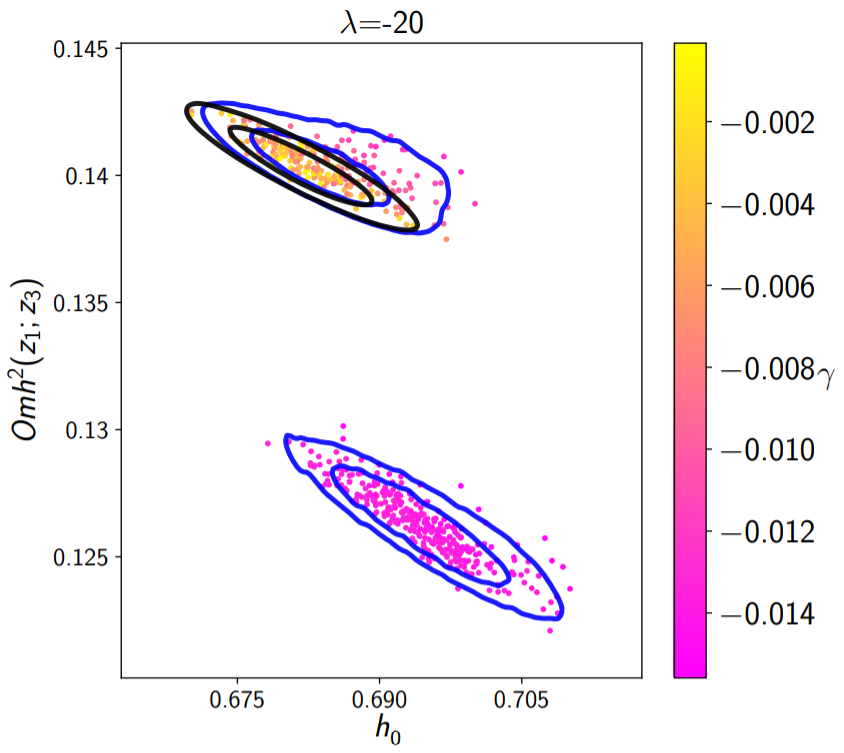}}%
\subfigure{\label{fig:b}\includegraphics[trim={0 0 0 0.90cm},clip, width=3.cm, height=4.cm]{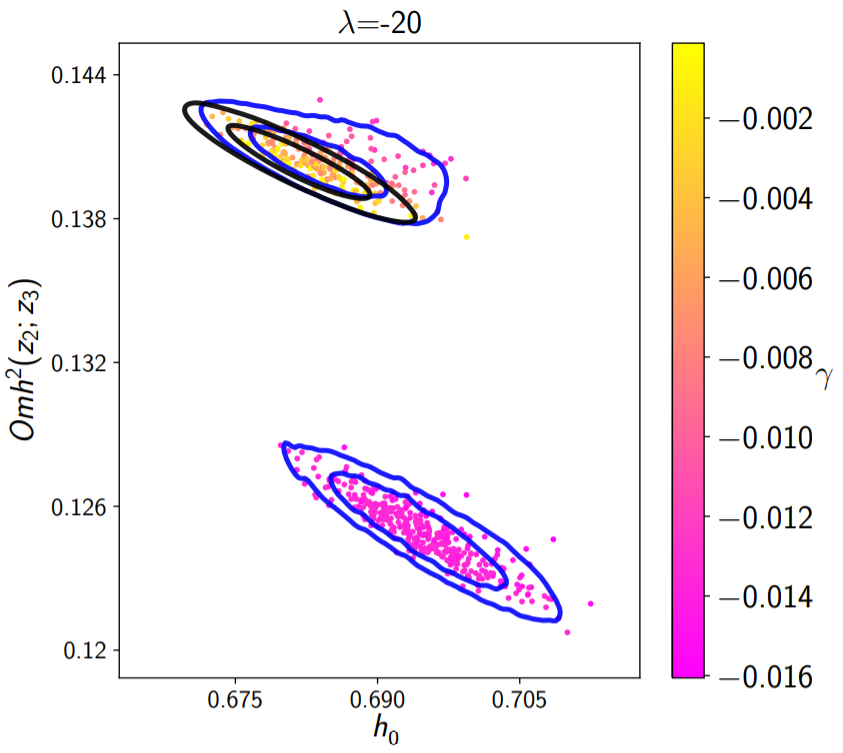}}
    \caption{$Omh^2$ diagnostic for the graduated dark energy model with $\lambda=-20$
    using three redshifts $\{z_1,z_2\}$ (left), $\{z_1,z_3\}$ (middle) and $\{z_2,z_3\}$ (right). The colour code indicates the value of $\gamma$ parameter, 
    where the yellow points mimic the $\Lambda$CDM behaviour and the pink ones
    the new feature introduced by the gDE model.}
    \label{fig:om_diag}
\end{figure}

We calculate, from \eqref{def:Omh2}, the model independent estimates as $Omh^2(z_{1};z_{2})=0.164 \pm 0.024$, $Omh^2(z_{1};z_{3})=0.123 \pm 0.006$ and $Omh^2(z_{2};z_{3})=0.119 \pm 0.007$ by using $H(z_{1}=0) = 69.8 \pm 0.8\, {\rm km\,s}^{-1}{\rm Mpc}^{-1}$ from the TRGB $H_0$ \cite{Freedman:2019jwv}, $H(z_{2}=0.57) = 97.9 \pm 3.4\,{\rm km\,s}^{-1}{\rm Mpc}^{-1}$ based on the clustering of galaxies in the SDSS-III BOSS DR11 \cite{Anderson:2013zyy}, and $H(z_{3}=2.34) = 222.4 \pm 5.0\,{\rm km\,s}^{-1}{\rm Mpc}^{-1}$ based on the BAO in the Ly-$\alpha$ forest of SDSS DR11 data \cite{Delubac:2014aqe}. We notice that the constraint $Omh^2 = 0.140 \pm 0.002$ ($Omh^2 =0.143 \pm 0.001$ in Planck 2018 \cite{Aghanim:2018eyx}) we obtained for the $\Lambda$CDM is in clear tension with the latter two of these estimates. We see in Table \ref{TableI} that, for $\lambda\leq-10$ as well as the $\lambda$ free case, the constraints for all of the three $Omh^2$ exhibit bimodal characteristic, i.e., there are two valued constraints  corresponding to the new (left) and old (right) maxima. We notice $Omh^2(z_{1};z_{2})\sim0.140$ (as in the $\Lambda$CDM) almost the same for both the new and old maxima, yet it agrees with the corresponding model independent estimate. However, when we consider $Omh^2(z_{1};z_{3})$ and $Omh^2(z_{2};z_{3})$ we observe that while the ones associated with the new maximum yield $\sim 0.125$ in agreement with the corresponding model independent estimates, the ones associated with the old maximum yield $\approx0.140$ in tension. For a visual demonstration, in Fig. \ref{fig:om_diag}, we show the marginalised posterior distributions for the parameter $\gamma$ in the $\{\gamma, Om  h^2(z_i ; z_j),  h_{0}\}$ subspace for $\{z_1,z_2\}$, $\{z_1,z_3\}$ and $\{z_2,z_3\}$, where the blue contours and 3D scatter color plots described the gDE-CDM model for $\lambda=-20$. The color code indicates the value of $\gamma$ labelled by the color bar. Black contours display 2D marginalised posterior distributions for the $\Lambda$CDM which agree with the position of the yellow points corresponding to the old maxima of the gDE-CDM. The contours at about $Omh^2\sim 0.125$ correspond to the new maxima of the gDE-CDM describing the case in which the energy density of the gDE passes below zero $z\sim 2.32$.

\begin{figure}[t!]
\captionsetup{justification=raggedright,singlelinecheck=false,font=footnotesize}
    \centering
     \includegraphics[trim = 1mm 1mm 1mm 0mm, clip, width=8.cm, height=4.cm]{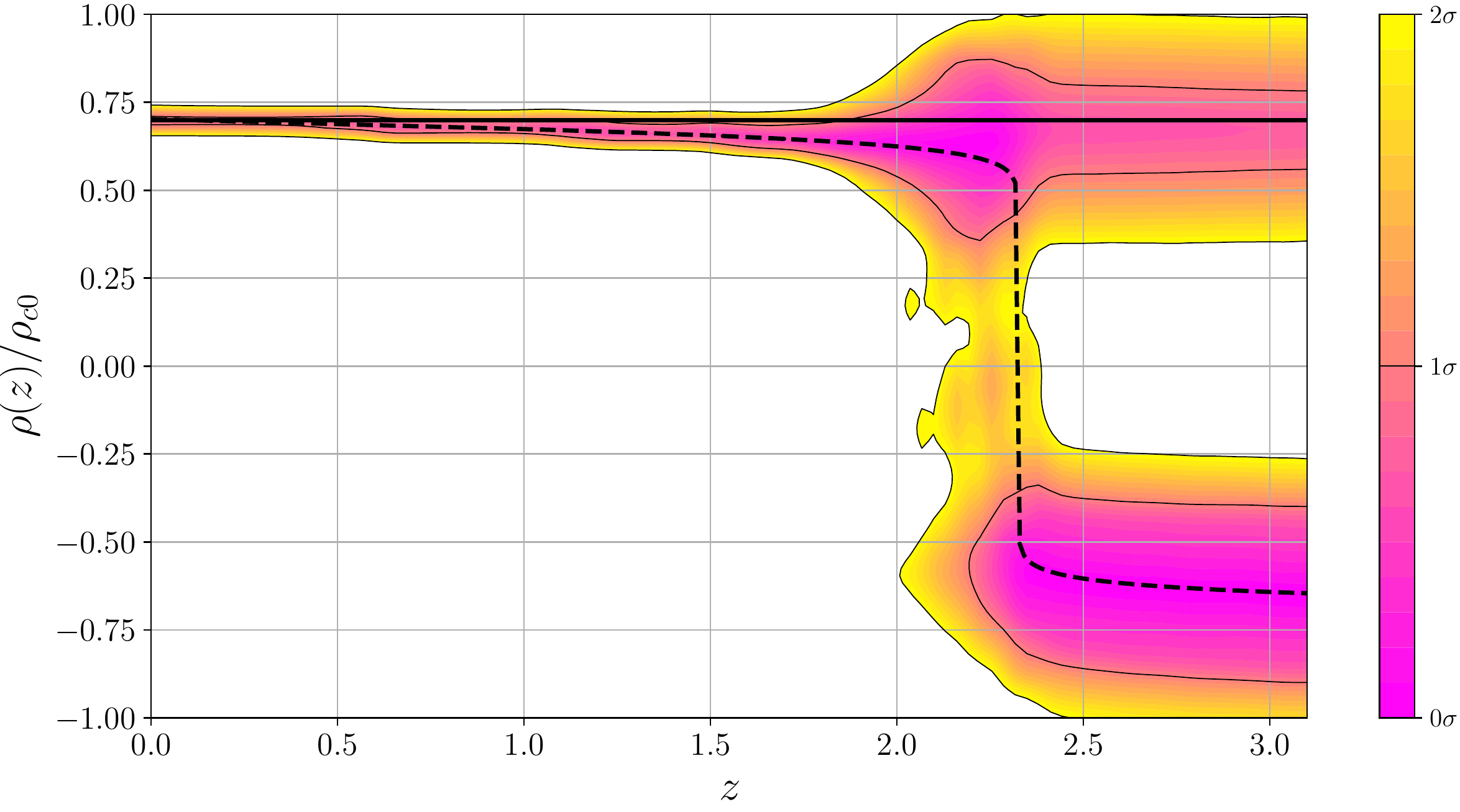}
    \includegraphics[trim = -6mm  1mm 1mm 0mm, clip, width=8.cm, height=4.cm]{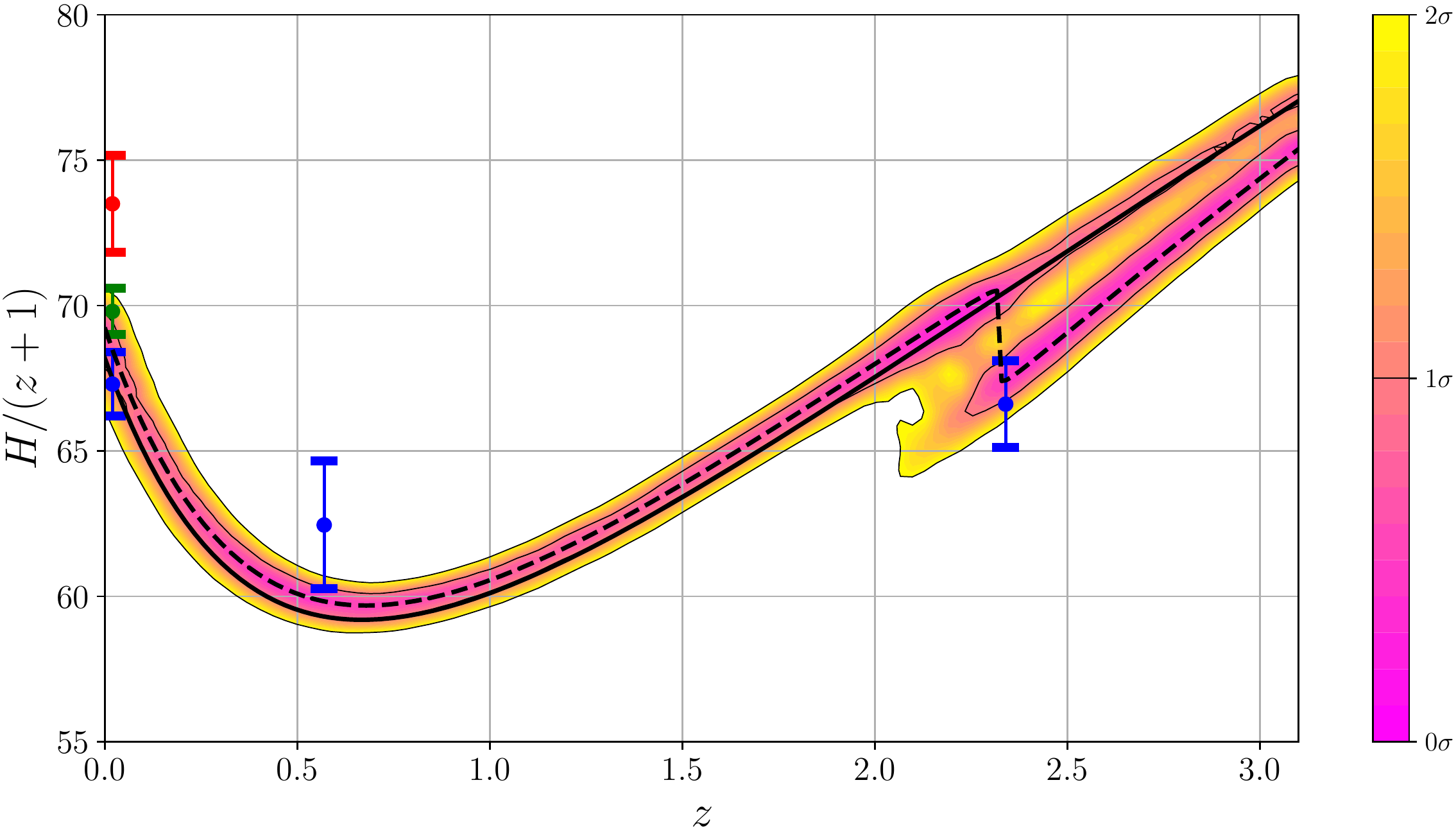}
    \caption{
    Top panel: $\rho_{\rm gDE}/\rho_{\rm c0}$ versus redshift $z$ for $\lambda=-20$ displays the maximum predicted that $\rho_{\rm gDE}$ changes sign at $z\sim2.3$. Bottom: $H(z)/(1+z)$ function. Include the latest BAO data points \cite{Aubourg:2014yra} (blue bars) where  $H_0=67.3\pm 1.1$, the Planck 2018 \cite{Aghanim:2018eyx} $H_0=67.4\pm 0.5$ data (red bar) and the TGRB model independent \cite{Freedman:2019jwv} $H_0=69.8\pm 0.8$ data (green bar). Black dashed line corresponds to best-fit values of gDE and solid black line corresponds to LCDM. We note that, due to the jump at $z\sim2.3$, the gDE model is not in tension with the BAO Ly-$\alpha$ data from $z=2.34$ in contrast to $\Lambda$CDM model and also gDE gives larger $H_0$ values w.r.t. $\Lambda$CDM model and thereby relaxes $H_0$ tension.}\label{fig:rhoandH}
\end{figure}

 All these superiorities in goodness of fit to the observational data arising in the case of the new maxima of the gDE-CDM are obviously consequences of the fact that the energy density of the gDE passes below zero at $z_*\approx2.3$ by exhibiting a certain type of dynamics. By using the {\it fgivenx} package \cite{fgivenxcite}, we show in the upper panel of Fig.~\ref{fig:rhoandH} the probability (the more pink implies more probable) distribution of the redshift dependency of the energy density of gDE scaled to the critical energy density of the present-day Universe, viz., $\rho_{\rm DE}/\rho_{\rm c,0}$. We observe that gDE, viz., $\rho_{\rm DE}(z)/\rho_{\rm c,0}$, does not distinguish from $\Lambda$ (solid straight black line) at a value $\sim0.70$ for $z\lesssim2$, but it reaches a junction at $z\sim2.3$ and for larger redshifts it either keeps tracking $\Lambda$ by retaining the value $\sim0.70$ (the one associated with the old maximum and disfavoured by the data) or rapidly changes route and starts to track a new value $\sim-0.70$ like a mirror image of the former track at $\rho_{\rm DE}=0$ (the case associated with the new maximum and favoured by the data). The rapid sign switch of the gDE energy density at $z\sim2.3$ implies a rapid drop in the total energy density of the Universe, and in $H(z)$, at that redshift. This behaviour of $H(z)$ emerges in association with the new maxima of the gDE-CDM for more negative values of $\lambda$, as can be seen in the lower panel of Fig.~\ref{fig:rhoandH}, reconciles it with the lower $H(z)$ value of the Ly-$\alpha$ data at $z=2.34$ with respect to the one predicted by $\Lambda$CDM for that redshift. Furthermore, this reconciliation between the gDE-CDM and Ly-$\alpha$ data, in turn, provides the gDE-CDM with easiness in achieving large $H(z)$ values for $z\lesssim 2$ and thereby predicts larger $H_0$, and so gDE-CDM relieves the $H_0$ tension that $\Lambda$CDM has been suffering from.

\section{Spontaneous sign switch in the cosmological constant}

In this section, we would like to continue by commenting on the implication of the dynamics of gDE that leads to all these reconciliations with the observational data on the nature of the dark energy. First, we note the following features of gDE that we have further understood upon confronting the observational data. We read off from Table \ref{TableI} that, for larger negative values of $\lambda$, $\rho_{\rm DE}/\rho_{\rm c,0}=0.70$ and $w_0\sim-1.01$ (i.e., in the phantom region but very close to the conventional vacuum energy) at $z=0$, its energy density switches sign rapidly (almost spontaneously) at $z_*\approx 2.32$ (which is quite stable) and settles into a value $\rho_{\rm DE}/\rho_{\rm c,0}\sim-0.70$ (the opposite of its present-day value) and remains ($w_{\rm DE}\approx-1$) there for $z_*\gtrsim 2.3$. Next, we observe in the same table that the larger the negative values of $\lambda$, the better fit to the data (the larger $\Delta\chi_{\rm min}^2$). This follows the trend that makes $\rho_{\rm DE}(z)$ increasingly resemble a step function centred at $z_*$ with two branches yielding opposite values about zero --a pattern of flat positive energy density for $z<z_*$ and flat negative energy density for $z>z_*$, both of which have the same absolute value-- and indeed,  we know from  \eqref{eqn:step}, that $\rho_{\rm DE}$ transforms into a step function for arbitrarily large negative values of $\lambda$. The largest negative $\lambda$ value we considered in our analyses is $-27$, yet it is easy to check mathematically that considering even larger negative values would not effect our results considerably since, for this value, the function $\rho_{\rm DE}(z)$ already closely resembles a step function. Thus, our results from the new maximum of the gDE for large negative values of $\lambda$ can safely be interpreted as the results one would obtain for a cosmological constant that achieved its present-day positive value by spontaneously switching sign at $z_*\sim2.3$, but was negative in the earlier stage of the universe.

Some general constraints that are typically applied to classical matter source, irrespective of its detailed description, may be utilised for further supporting our interpretation (see \cite{EllisRC,Carroll:2003st}). Let us consider gDE as an actual barotropic fluid, $p=p(\rho)$, along with the best fit values obtained on its free parameters from the observational analysis. In this case, although it behaves almost like a cosmological constant (in spite of the fact that it switches sign at $z\approx2.32$) throughout the history of the universe, strictly speaking, it violates the weak energy condition, namely, the non-negativity conditions on the energy density, $\rho\geq0$, for $z>z_*$, and on the inertial mass density, $\rho_{\rm inert}\geq0$, throughout the history of the universe. Moreover, there are periods during which it violates the condition $0\leq c_s^2\leq1$ on the speed of sound of a barotropic fluid given by the adiabatic formula $c_s^2={\rm d}p/{\rm d}\rho$. The upper limit (causality limit) is a rigorous one which cannot be violated unless we abandon relativity theory. The lower limit applies to a stable situation, and otherwise the fluid is classically unstable against small perturbations of its background energy density -the so called Laplacian (or gradient) instability. It is well known that phenomenological fluid models of DE are difficult to motivate, and adiabatic fluid models are typically unstable against perturbations, since $c_s^2$ is usually negative for $w<0$. It is possible to evade this constraint in non-adiabatic fluid descriptions (e.g., canonical scalar field for which the effective speed of sound --which governs the growth of inhomogeneities in the fluid-- is equal to unity, $c_{\rm s\, eff}=1$), and in an adiabatic fluid if $w$ decreases sufficiently fast as the universe expands (e.g., Chaplygin gas). However, with some exceptions, it is unlikely to describe gDE with a canonical scalar field ---especially when we consider the best fit values. Also, gDE yields $c_s^2=-1+\gamma\lambda \left(\frac{\rho}{\rho_0}\right)^{\lambda-1}=-1+\frac{\gamma\lambda }{1+3\gamma (\lambda-1) \ln a}$, and $c_s^2(z=0)=-1+\gamma\lambda$. Accordingly, the constrains we obtained when $\lambda$ is free predict $c_s^2(z=0)=-0.6957\pm0.1739$ for $z=0$ and $c_s^2\gg1$ while $0<\rho\ll \rho_0$ (just after gDE assumes positive values at $z\approx2.32$). On the other hand, whether it is positive or negative, a cosmological constant (viz., the limit $\lambda\rightarrow-\infty$, see \eqref{eqn:step}) is well behaved: $\rho_{\rm inert}=0$, and $c_s^2=0$ (it has no speed of sound, and thereby does not support classical fluctuations). Regarding the negativity of its energy density (when $z>z_*$), a negative cosmological constant is ubiquitous in the fundamental theoretical physics without any complication, for instance, it can be taken as just a geometrical component ($\rho<0$ will then be an effective energy density rather than an actual one), and it also is very natural from symmetry considerations and provides the ground state (AdS background) in various low energy limits of string theory.

Thus, bringing all these points together, it is tempting to conclude that the cosmological constant has spontaneously switched sign and become positive at $z\approx 2.32$ and triggered the late-time acceleration. Of course, one could look for realising such a nontrivial behaviour of gDE as an effective source in a modified gravity model (the general constraints that are typically applied to classical matter source might then be evaded) and reach different conclusions.

\section{Conclusions}

We have considered a type of dark energy that can be viewed as characterising the minimum dynamical deviation from the null inertial mass density --described by the conventional vacuum (or cosmological constant, $\Lambda$)-- in the form $\rho_{\rm inert}\propto \rho^{\lambda}$ with $\lambda$ being a constant. This source, we called \textit{graduated Dark Energy} (gDE), presents a wide variety of dynamics which were first studied in the context of inflaton \cite{Barrow:1990vx,Barrow:1990nv,Barrow:1990td} and more recently of dark energy \cite{Nojiri:2004pf,Stefancic:2004kb,Stefancic:2005cs,Frampton:2011sp}. We focused on its dynamics (which has not been studied in detail so far) that emerges when $\rho_{\rm inert}<0$, and $\lambda<1$ is written as a ratio of two odd integers. In this case it yields an energy density that dynamically assumes negative values in the recent past, in line, for instance, with \cite{Aubourg:2014yra,Sahni:2014ooa,Delubac:2014aqe,Mortsell:2018mfj,Poulin:2018zxs,Capozziello:2018jya,Wang:2018fng,Dutta:2018vmq,Visinelli:2019qqu}. They proposed such models to address, for instance, the persistent tensions arising between the cosmological constant hypothesis of the standard $\Lambda$CDM model and the model independent $H_0$ measurements and/or high precision Ly-$\alpha$ measurements of BAO. Importantly, for large negative values of $\lambda$, gDE presents a phenomenological model described by a smooth function. It approximately describes the cosmological constant spontaneously switching sign at a certain redshift to become positive quite recently in the late universe. In particular, it transforms into a step function for arbitrarily large negative $\lambda$ values.

We have confronted the gDE-CDM model, replaced the $\Lambda$ hypothesis by the gDE, with the latest combined observational data sets of PLK+BAO+SN+$H$. We have observed that something striking occurs when $\lambda\leq-4$ (also when $\lambda$ is free): that the data predicts bimodal posterior probability distributions for the parameters of the gDE-CDM model: new maxima significantly far away from $\gamma=0$ ($\Lambda$CDM), and old maxima containing $\gamma=0$. The improvement in the goodness of the fit with respect to the $\Lambda$ reaches highly significant levels --e.g., $\Delta\chi_{\rm min}^2=6$ when $\lambda=-20$ and $\Delta\chi_{\rm min}^2=6.4$ when $\lambda$ is free-- for the new maxima, while it remains always at insignificant levels --$\Delta\chi_{\rm min}^2\lesssim0.02$, irrespective of the value of $\lambda$-- for the old maxima. We have shown that, in contrast to the old maxima covering the $\Lambda$CDM model, these new maxima of the gDE-CDM model also agree with the model-independent $H_0$ measurements, high-precision Ly-$\alpha$ data, and model-independent $Omh^2$ diagnostic estimates.

We have demonstrated that the superior features endowed by the new maxima of the gDE-CDM model are due to the energy density of the gDE rapidly changing sign at the redshift $z\approx2.3$ (shown to be quite stable in our observational analysis) and this in turn leads to a rapid drop in the total energy density of the universe, and in $H(z)$, at the same redshift. It has turned out that this happens for large negative values of $\lambda$, which renders the redshift dependency of the gDE density close to a step function, which to a good approximation describes a cosmological constant spontaneously  switching sign. Therefore, our findings, by means of gDE in the light of observational data, provide strong hints of a spontaneous sign switch in the cosmological constant. This leads us to conjecture that the cosmological constant has spontaneously switched sign and became positive, namely, the universe has transitioned from AdS vacua to dS vacua, at $z\approx 2.32$ and triggered the late-time acceleration. This  suggests looking for such mechanisms in string theory constructions. The fact that constructing metastable dS and/or AdS in string theory occupy a key place in the string theory investigations, indicates that the future confirmation or falsification of our conjecture would have far reaching implications for fundamental theoretical physics as well as for the identity of the dark energy.

\begin{acknowledgements}
The authors thank to Dragan Huterer, Paolo Creminelli, Jorge Nore\~{n}a and Mehmet Ozkan for valuable discussions. \"{O}.A. acknowledges the support by the Turkish Academy of Sciences in scheme of the Outstanding Young Scientist Award  (T\"{U}BA-GEB\.{I}P). \"{O}.A. is grateful for the hospitality of the Abdus Salam International Center for Theoretical Physics (ICTP) while the part of this research was being carried out. J.D.B. was supported by the STFC of the UK.
J.A.V. acknowledges the support provided by FOSEC SEP-CONACYT Investigaci\'on B\'asica A1-S-21925, and UNAM-DGAPA-PAPIIT  IA102219.
\end{acknowledgements}


\begin{thebibliography}{99}
\bibitem{ourfootnote}
  The mathematical expressions of the form $x^y$, when $x<0$ and $y$ is not an integer, should be treated carefully. The correct result of $x^y$ is mathematically well-defined without ambiguity, if it is treated with infinite numeric precision. When $x<0$, then $x^y$ is real-valued exactly when $y$ can be written as a fraction, $m/n$, where $m$ is an integer and $n$ is an odd integer. Furthermore, the result is positive when $m$ is even and negative when $m$ is odd. When $y$ cannot be written as such, the result would be an imaginary number. Hence, to get $\rho<0$ for $a<a_*<1$ (or $z>z_*>0$), we must choose $\lambda<1$ values in accordance with these rules. The lack of continuity for $\lambda$ arising from this situation would also present a practical problem for the computer models that want to raise negative numbers to fractional powers. Because computers do store fractional numbers employing finite-precision representations, the value stored in memory should always be considered a close approximation to the true value that is being represented. When a negative number is raised to a fractional power, there is no way to know whether that true value is even rational, and if it is, whether it would correspond to the positive or to the negative solution. To get around this issue when we investigate gDE computationally under the conditions given in \eqref{1stconditions} we write $\frac{\rho}{\rho_0}=x^y$ in an equivalent way as $\frac{\rho}{\rho_0}={\rm sgn}(x)\, {|x|}^y$ for $y=\frac{m}{n}$ with $m$ and $n$ being odd integers.
\end{thebibliography}
\end{document}